\documentclass[%
 reprint,
 amsmath,amssymb,
 aps,
]{revtex4-2}

\usepackage{graphicx}
\usepackage{dcolumn}
\usepackage{bm}

\usepackage{braket}
\usepackage{xparse}
\usepackage{siunitx}

\newcommand{\bx}{\mathbf{x}}

\newcommand{\hilbert}{\mathcal{H}}
\newcommand{\THawk}{T_{\text{H}}}
\newcommand{\RS}{R_{\text{S}}}
\newcommand{\taumid}{\tau_{\text{mid}}}
\newcommand{\rmid}{r_{\text{mid}}}

\newcommand{\response}{\mathcal{F}}
\newcommand{\responseB}{\response_\text{B}}
\newcommand{\responseH}{\response_\text{H}}
\newcommand{\responseU}{\response_\text{U}}

\newcommand{\FB}{F_\text{B}}
\newcommand{\FH}{F_\text{H}}

\newcommand{\vacM}{\ket{0_\text{M}}}

\newcommand{\teff}{T_\mathrm{eff}}
\newcommand{\tKMS}{T_\mathrm{KMS}}

\DeclareMathOperator{\sinc}{sinc}
\newcommand{\defeq}{\equiv}
\newcommand{\del}{\partial}
\newcommand{\brac}[2]{ \left( \frac{#1}{#2} \right) } 

\NewDocumentCommand{\dd}{o m g g }{\frac{d\IfValueT{#1}{^#1}\IfValueT{#3}{#2}}{d\IfValueTF{#3}{#3}{#2}\IfValueT{#1}{^#1}\IfValueT{#4}{\d #4}}}
\NewDocumentCommand{\deldel}{o m g g }{\frac{\del\IfValueT{#1}{^#1}\IfValueT{#4}{^2}{\IfValueT{#3}{#2}}}{\del\IfValueTF{#3}{#3}{#2}\IfValueT{#1}{^#1}\IfValueT{#4}{\del #4}}}

\def\vk{{\vec{k}}}

\def\vv{{\vec{v}}}

\def\vx{{\vec{x}}}

\usepackage{enumerate}

\usepackage{xcolor}

\newcommand{\red}{\color{black}}

\begin{document}

\preprint{APS/123-QED}

\title{What Hawking Radiation Looks Like as You Fall into a Black Hole}

\author{Christopher J. Shallue}
 \email{cshallue@cfa.harvard.edu}
\affiliation{%
Center for Astrophysics \textbar\ Harvard \& Smithsonian, Cambridge, MA 02138
}%

\author{Sean M. Carroll}
\affiliation{
  Departments of Physics \& Astronomy and Philosophy, Johns Hopkins University,
  Baltimore MD 21218
  and
  Santa Fe Institute, Santa Fe, NM 87501 \vspace{8pt}
}%

\date{\today}

\begin{abstract}
  We study the measurements of a freely falling Unruh-DeWitt particle detector near the horizon of a semiclassical Schwarzschild black hole. Our results show that the detector's response increases smoothly as it approaches and crosses the horizon in both the Hartle-Hawking and Unruh vacua. However, these measurements are dominated by the effects of switching the detector on and off, rather than by the detection of Hawking radiation particles. We demonstrate that a freely falling Unruh-DeWitt detector cannot directly measure Hawking radiation near the horizon because the time required for thermalization is longer than the time spent near the horizon. We propose an operational definition of the effective temperature along an infalling trajectory based on measurements by a particle detector. Using this method, we find that the effective temperature measured by a freely falling observer in the Hartle-Hawking vacuum increases smoothly from the Hawking temperature far from the horizon to twice the Hawking temperature at the horizon, and continues to rise into the interior of the black hole. This effective temperature closely matches an analytical prediction derived by embedding Schwarzschild spacetime into a higher-dimensional Minkowski space, suggesting that further exploration of higher-dimensional embeddings could provide new insights into the near-horizon behavior of black holes.
\end{abstract}

\maketitle


\section{Introduction}\label{sec:intro}

In his landmark 1975 derivation, Hawking showed that observers located far from a black hole, at late times after its formation, will measure a thermal spectrum of particles emanating from the black hole at what is now called the Hawking temperature~\cite{hawking_ParticleCreationBlack_1975}.
This result was obtained using a semiclassical approximation that combines quantum field theory with classical general relativity, while neglecting the back-reaction of the field onto the spacetime.
Nonetheless, its predictions are expected to remain valid in a more fundamental theory of quantum gravity, as it focuses on asymptotically flat regions---those far from the black hole, both long before and long after its formation---where the semiclassical approximation is expected to apply.

Although Hawking radiation is well understood in these asymptotic regions, significant questions remain about its nature closer to the event horizon, even at late times after the black hole has settled into a semistationary state. 
Where and how is Hawking radiation generated? {\red A traditional heuristic story says} that particle pairs are created just outside the horizon, with one particle falling into the black hole and the other escaping to infinity~\cite{hawking_ParticleCreationBlack_1975,wald_ParticleCreationBlack_1975,davies_EnergymomentumTensorEvaporating_1976,hartle_PathintegralDerivationBlackhole_1976,israel_ThermofieldDynamicsBlack_1976,parikh_HawkingRadiationTunneling_2000}. Others argue that Hawking radiation is produced nonlocally much further from the horizon~\cite{giddings_HawkingRadiationStefan_2016,bardeen_SemiclassicalStressenergyTensor_2017,nikolic_SemiclassicalSolutionBlack_2024}.
{\red Additionally, motivated by the ongoing debate about information loss in black hole evaporation~\cite{hawking_BreakdownPredictabilityGravitational_1976,mathur_InformationParadoxPedagogical_2009,pegas_GloballyHyperbolicEvaporating_2025}, there has been recent speculation about the quantum-gravitational nature of the horizon, including the conjecture of a high-energy ``firewall'' in its vicinity~\cite{almheiri_BlackHolesComplementarity_2013}.}

In this paper we treat the black hole semiclassically, putting aside full quantum gravity and firewalls, and study what an observer would experience if they fell into a semiclassical black hole. There is an apparent tension between two accepted pieces of Hawking-radiation lore: inertial observers far away see the black hole as a thermal object with definite temperature, while observers falling across the horizon are supposed to see a locally Minkowski vacuum state, in accordance with the equivalence principle~\cite{unruh_NotesBlackholeEvaporation_1976,crispino_UnruhEffectIts_2008}.
But as an observer falls in from infinity, outgoing radiation presumably blueshifts in their reference frame. How can both {\red ideas} be true? Does Hawking radiation effectively turn off at some point?

Crucially, freely falling observers only have a limited amount of time to make measurements near the horizon. An observer falling from rest at infinity spends proper time of order $\RS$ near the horizon, where $\RS$ is the Schwarzschild radius. Is this enough time to measure Hawking radiation? We will show that the answer is \textit{no}, at least for an observer carrying an Unruh-DeWitt detector~\cite{unruh_NotesBlackholeEvaporation_1976,dewitt_QuantumGravityNew_1979} coupled to a massless scalar field in Schwarzschild spacetime, {\red because their detector does not} have enough time to thermalize in the near-horizon region (also see Ref.~\cite{kaplanek_QubitsHorizonDecoherence_2021}). {\red Therefore, their measurements will always be dominated by the effect of switching the detector on and off, rather than by Hawking radiation.} 
This reconciles the apparent tension noted above: an infalling observer sees blueshifted radiation along their path, but very close to the horizon does not have enough time to detect any substantial radiation at all.
However, we will demonstrate that the observer can still measure an effective local temperature, even across the horizon and into the interior of the black hole.

It is well known that \textit{static} observers outside a Schwarzschild black hole in the Hartle-Hawking vacuum~\cite{hartle_PathintegralDerivationBlackhole_1976,israel_ThermofieldDynamicsBlack_1976} will detect a thermal spectrum of particles, with the local temperature increasing the closer they are to the horizon~\cite{unruh_NotesBlackholeEvaporation_1976,candelas_VacuumPolarizationSchwarzschild_1980}. This is at least partly caused by the acceleration needed to keep these observers at a fixed distance from the black hole. Even in the Minkowski vacuum, uniform acceleration causes a detector to register thermal particles, a phenomenon known as the Unruh effect~\cite{unruh_NotesBlackholeEvaporation_1976}. In order to remove the confounding effects of acceleration, it is desirable to study inertial observers. References~\cite{smerlak_NewPerspectivesHawking_2013,hodgkinson_StaticDetectorsCirculargeodesic_2014} studied observers on circular geodesics around Schwarzschild black holes, finding that they measure approximately thermal radiation at temperatures greater than static detectors at the same radii. However, bound circular orbits only exist for ${r > 1.5 \RS}$, where $r$ is the Schwarzschild radial coordinate, so this class of observers cannot shed light on the region closer to the horizon. Recently, Ref.~\cite{ng_LittleExcitementHorizon_2022} presented the first detailed calculation of measurements by a freely falling Unruh-DeWitt detector near the horizon of a 4-dimensional Schwarzschild black hole. Surprisingly, they found the detector measured a nonmonotonic response, with a local maximum appearing near the horizon. Similar findings have been reported for a detector falling into a 3-dimensional Ba\~nados-Teitelboim-Zanelli (BTZ) black hole~\cite{preciado-rivas_MoreExcitementHorizon_2024,wang_SingularExcitementHorizon_2024a}, but not a 2-dimensional Schwarzschild black hole analog~\cite{juarez-aubry_OnsetDecay1+1_2014}. We study the 4-dimensional Schwarzschild case in detail, finding that a freely falling detector measures a monotonically increasing response across the horizon, in contrast to the findings of Ref.~\cite{ng_LittleExcitementHorizon_2022}. We also study how to interpret and utilize these measurements, since, as previously mentioned, the detector's response near the horizon is dominated by switching effects and is not a direct measure of Hawking radiation.

Freely falling observers have also been studied using approximate analytical methods, often by calculating local temperature functions along their trajectories. The methods used and conclusions drawn vary significantly. Reference~\cite{massar_EnergymomentumTensorEvaporating_1993} derived a local temperature for observers falling into a stationary 2-dimensional black hole, finding that the temperature vanishes on the horizon. Meanwhile, Refs.~\cite{deeg_QuantumAspectsBlack_2006,greenwood_HawkingRadiationSeen_2009,
barbado_HawkingRadiationPerceived_2011,barbado_HawkingRadiationPerceived_2012,smerlak_NewPerspectivesHawking_2013,singh_BlackHoleKinematics_2014,chakraborty_QuantumPeekBlack_2015} considered stationary black holes as well as those formed by gravitational collapse, finding that freely falling observers experience finite, positive temperatures at the horizon, dependent on the velocity of the observer.
Local temperature functions have also been derived by extending the classical Tolman gradient~\cite{tolman_WeightHeatThermal_1930,tolman_TemperatureEquilibriumStatic_1930,santiago_TolmanlikeTemperatureGradients_2018} to black-hole spacetimes~\cite{gim_QuantalTolmanTemperature_2015,eune_EffectiveTolmanTemperature_2017,kim_EffectiveTolmanTemperature_2017}.
Finally, Ref.~\cite{brynjolfsson_TakingTemperatureBlack_2008} derived a local temperature function by globally embedding 4-dimensional Schwarzschild spacetime into 6-dimensional Minkowski spacetime~\cite{fronsdal_CompletionEmbeddingSchwarzschild_1959}. The authors used the 6-acceleration $a$ of an observer instantaneously at rest on a radial geodesic in the Schwarzschild spacetime to compute a local Unruh temperature~\cite{unruh_NotesBlackholeEvaporation_1976},
\begin{equation}\label{eq:T_brynjolfsson}
  T = \frac{a}{2\pi} =\THawk \sqrt{1 + \brac{2M}{r} + \brac{2M}{r}^2 + \brac{2M}{r}^3},
\end{equation}
where $M$ is the mass of the black hole and $\THawk = 1 / (8 \pi M)$ is its Hawking temperature.
This technique has previously been used to produce the expected local temperature for a variety of observers in different black hole spacetimes~\cite[e.g.,][]{deser_EquivalenceHawkingUnruh_1998,deser_MappingHawkingUnruh_1999,chen_GEMSApproachStationary_2004,santos_GlobalEmbeddingDdimensional_2004,langlois_CausalParticleDetectors_2006,banerjee_NewGlobalEmbedding_2010}, despite the fact that detailed measurements by particle detectors differ between the ambient and embedded spacetimes due to the dimension-dependence of the Unruh effect~\cite{takagi_VacuumNoiseStress_1986,unruh_AcceleratedMonopoleDetector_1986,langlois_ImprintsSpacetimeTopology_2005}.

Despite the variety of local temperature functions proposed to describe the experiences of infalling observers, none have been compared to detailed calculations of particle detector measurements in 4-dimensional spacetimes. Doing so is important for several reasons. First, given the range of assumptions and differing predictions in these approximations, it would be valuable to determine which {\red are supported by detailed calculations, thereby validating or refuting the underlying assumptions in each case}. Second, the ``temperatures'' calculated are not expected to be exact, as the nonstationary trajectory of an infalling observer precludes true equilibrium. Understanding how closely the observer perceives a thermal state would provide valuable insights into the near-horizon region and clarify the accuracy of these analytical approximations.

In this work, we propose using an Unruh-DeWitt detector to measure the effective local temperature along the trajectory of a freely falling observer, even when the detector's measurements are dominated by switching effects. We find that the temperature in the Hartle-Hawking vacuum aligns closely with Eq.~\eqref{eq:T_brynjolfsson} across a range of detector energies and measurement locations, even within the black hole's interior.
Our results therefore support the global embedding method for calculating the local temperature of Hawking radiation, even on nonstationary trajectories where some authors have expressed skepticism about its viability~\cite{chen_NoteGeneralizationGlobal_2005,hodgkinson_HowOftenDoes_2012}.

This paper is organized as follows. In Section~\ref{sec:detectors}, we review the Unruh-DeWitt particle detector model used throughout the paper. In Section~\ref{sec:minkowski}, we study inertial detectors in thermal states in Minkowski space, where we obtain results and develop intuition relevant to the black hole case. In Section~\ref{sec:schwarz}, we describe the formalism and numerical methods used to calculate measurements by particle detectors in Schwarzschild spacetime. We present our results in Section~\ref{sec:results}, where we also define and study the effective temperature along an infalling trajectory. We conclude in Section~\ref{sec:conclusions} with a summary of our findings.

Throughout this paper we use units in which $c = G = \hbar = k_\text{B} = 1$. In 4-dimensional spacetimes we use the metric signature ${(+,-,-,-)}$. $\hat{f}(\omega)$ denotes the Fourier transform of $f(t)$,
\begin{equation}
  \hat{f}(\omega) \defeq \int_{-\infty}^\infty dt \, e^{- i \omega t} f(t).
\end{equation}

\section{Particle detectors}\label{sec:detectors}

\subsection{Setup}

We begin by reviewing the Unruh-DeWitt particle detector model~\cite{unruh_NotesBlackholeEvaporation_1976,dewitt_QuantumGravityNew_1979} that we use in this paper. For more detailed reviews, see Refs.~\cite{birrell_QuantumFieldsCurved_1982,takagi_VacuumNoiseStress_1986,langlois_ImprintsSpacetimeTopology_2005,louko_TransitionRateUnruhDeWitt_2008,hodgkinson_ParticleDetectorsCurved_2013}.

Our detector is an idealized point particle with internal structure described by Hamiltonian $H_\mathrm{D}$ acting on the Hilbert space $\hilbert_\mathrm{D}$. It has two energy eigenstates, $\ket{0}$ and $\ket{E}$, corresponding to eigenvalues $0$ and $E$. $E$ is called the detector's \textit{energy gap}. It is endowed with a so-called ``monopole moment'' operator~\cite{dewitt_QuantumGravityNew_1979,takagi_VacuumNoiseStress_1986}, whose Heisenberg-picture dynamics are given by 
\begin{equation}
  \mu(\tau) = e^{i H_\mathrm{D} \tau} \mu(0) e^{-i H_\mathrm{D} \tau}.
\end{equation}

The detector is linearly coupled to a real scalar field $\phi(x)$ via the interaction Hamiltonian
\begin{equation}\label{eq:H_int}
  H_\text{int}(\tau) = \lambda \chi(\tau) \mu(\tau) \phi(\bx(\tau))
\end{equation}
acting on the joint Hilbert space $\hilbert_\mathrm{\phi} \otimes \hilbert_\mathrm{D}$, where $\hilbert_\mathrm{\phi}$ is the Hilbert space of the field. Here, $\bx(\tau)$ denotes the trajectory of the detector, $\tau$ is its proper time, $\lambda$ is a small coupling parameter, and $\chi$ is a real nonnegative \textit{switching function} that allows the coupling to be time-dependent. $\chi$ is required to be smooth and vanish at early and late times on the trajectory~\cite{satz_ThenAgainHow_2007,louko_TransitionRateUnruhDeWitt_2008}. We assume the observer carrying the detector is free to choose $\chi$ and therefore control where in the trajectory the detector is active.

More general detector models have been studied in various contexts, such as detectors with finite extent and detectors coupled nonlinearly to the field or to its derivatives. See Ref.~\cite{langlois_ImprintsSpacetimeTopology_2005} for a review.

If the initial ($\tau \to -\infty$) state of the field and detector is $\ket{\Psi} \otimes \ket{0}$, then to first order in the coupling parameter $\lambda$, the probability that the final ($\tau \to \infty$) state of the detector is $\ket{E}$ is given by
\begin{equation}\label{eq:P(E)}
  P(E) = \lambda^2 |\braket{E | \mu(0) | 0} |^2 \, \response(E),
\end{equation}
where
\begin{equation}\label{eq:F(E)-def}
  \response(E) \defeq \int_{-\infty}^{\infty} d\tau \, \int_{-\infty}^{\infty} d\tau' \,e^{-i E(\tau - \tau')} \chi(\tau) \chi(\tau') W(\tau, \tau'),
\end{equation}
and where $W(\tau,\tau') \defeq W(\bx(\tau), \bx(\tau'))$ is the pull-back of the positive-frequency Wightman distribution
\begin{equation}\label{eq:W-def}
  W(x, x') \defeq \braket{\Psi | \phi(x) \phi(x') | \Psi}
\end{equation}
to the detector's worldline. In a minor abuse of notation, we use $W(x,x')$ to denote the Wightman distribution and $W(\tau, \tau')$ to denote its pullback, since the meaning will be clear from the context.

$W(x,x')$ is a distribution, so the integral in Eq.~\eqref{eq:F(E)-def} is formally a distributional integral. One way to calculate this integral is to represent $W$ as a family of functions $W_\epsilon$, where $\epsilon > 0$ is a regularization parameter, and take the limit as $\epsilon \to 0$ after computing the integral~\cite{birrell_QuantumFieldsCurved_1982,schlicht_ConsiderationsUnruhEffect_2004}. References~\cite{satz_ThenAgainHow_2007,louko_TransitionRateUnruhDeWitt_2008} have calculated explicit expressions for this limit in 4-dimensional spacetimes. In this paper, we will sidestep these complications by representing $W$ as an integral expression over ordinary functions (e.g., as a Fourier transform). Then by reordering the integrals in Eq.~\eqref{eq:F(E)-def}, we will express $\response$ as an ordinary integral, {\red thus avoiding the distributional nature of $W$ in our numerical calculations}.

If the detector's energy gap is positive ($E>0$), then $\ket{0}$ is the ground state and $\ket{E}$ is an excited state. In this case, a transition from $\ket{0}$ to $\ket{E}$ can be interpreted as the detector absorbing a particle with energy $E$ from the field. $P(E)$ is therefore the probability that a particle will be detected, or, equivalently, the fraction of detectors in an ensemble of identical detectors that will detect a particle. However, care must be taken with this interpretation if the detector is switched on and off too rapidly, because this can also cause the detector to be excited, even if the field contains no particles~\cite{unruh_WhatHappensWhen_1984,birrell_QuantumFieldsCurved_1982,sriramkumar_ResponseFinitetimeParticle_1996}. We will study the effects of switching in detail in Section~\ref{sec:minkowski}.

On the other hand, if the energy gap is negative, a transition from $\ket{0}$ to $\ket{E}$ can be interpreted as the detector emitting a particle with energy $E$, where $P(E)$ is the probability of an emission.
The detector's response is said to be thermal if it satisfies the \textit{detailed balance}~\cite{takagi_VacuumNoiseStress_1986} form of the Kubo-Martin-Schwinger (KMS) condition~\cite[][]{kubo_StatisticalMechanicalTheory_1957,martin_TheoryManyParticle_1959},
\begin{equation}\label{eq:detailed-balance}
  P(E) = e^{-E/\tKMS} P(-E),
\end{equation}
where $\tKMS$ is the KMS temperature. This condition expresses thermal equilibrium between emission and absorption.

Since our goal is to measure Hawking radiation, we mainly focus on detector excitations ($E>0$) rather than de-excitations. Consequently, we will generally refer to transitions as ``detecting'' particles.  However, in cases where $E<0$, it should be understood that transitions indicate particle emissions rather than detections.

The function $\response(E)$ is called the detector's \textit{response function}. It is standard practice to focus on $\response$ instead of $P$, because $\response$ depends only on the detector's trajectory and the state of the field, whereas the coefficient $\lambda^2 |\braket{E | \mu(0) | 0} |^2$ depends only on the internal details of the detector.
In an ensemble of identical detectors, $\response$ is proportional to the expected number of detector transitions. Throughout this paper, we will informally refer to a single detector ``measuring''  $\response$, with the understanding that this actually requires an ensemble of detectors.
We emphasize that $\response$ is {\red calculated from an integral over the entire trajectory} and is thus localized only within the region where $\chi$ is nonvanishing. In other words, while a nonzero value of $\response$ can indicate that particles were detected, it does not give precise information about \textit{where} those particles were detected.

In some circumstances, it is useful to define a \textit{response rate} $\dot\response$. On stationary trajectories (see Section~\ref{sec:stationary}), $\dot\response$ can be defined as the limit of $\response / \Delta\tau$ as $\Delta\tau \to \infty$, where $\Delta\tau$ is the proper duration of the measurement, in which case $\dot\response$ is the average rate of particles detected along the trajectory~\cite{birrell_QuantumFieldsCurved_1982}. However, on nonstationary trajectories, such as a detector falling into a black hole, defining $\dot\response$ is more subtle and requires taking a limit as the detector switch-on and switch-off profiles become arbitrarily sharp~\cite{satz_ThenAgainHow_2007,louko_TransitionRateUnruhDeWitt_2008}. An advantage of $\dot\response$ is that it does not require explicitly choosing a switching function $\chi$. However, like $\response$, it is integrated over the entire region of the trajectory where the detector is switched on, so it does not give a truly local measurement. Moreover, it is difficult to relate $\dot\response$ to the actual experience of an observer, for it does not say anything about the response of a single detector (or ensemble of detectors), rather it compares two separate ensembles of detectors~\cite{langlois_ImprintsSpacetimeTopology_2005,louko_TransitionRateUnruhDeWitt_2008}. Finally, for a detector falling into a black hole, $\dot\response$ is much more numerically challenging to compute than $\response$. We attempted to compute $\dot\response$ using the method described in Ref.~\cite{hodgkinson_ParticleDetectorsCurved_2013}, but were unable to get the calculations to converge.

\subsection{Switching Function}

{\red The integral in Eq.~\eqref{eq:F(E)-def} is guaranteed to converge if the switching function $\chi$ is smooth and of compact support~\cite{kay_TheoremsUniquenessThermal_1991,satz_ThenAgainHow_2007,louko_TransitionRateUnruhDeWitt_2008}. For Rindler observers, Ref.~\cite{higuchi_UniformlyAcceleratedFinitetime_1993} found that it is sufficient for $\chi$ to be continuous. 
For freely falling detectors in the Schwarzschild spacetime, Ref.~\cite{ng_LittleExcitementHorizon_2022} found the following switching function sufficiently regular for numerical convergence:}
\begin{equation}\label{eq:chi-cos}
  \chi(\tau) =
  \begin{cases}
    \cos^4\left(\dfrac{\pi(\tau - \taumid)}{2 \Delta\tau}\right), & |\tau - \taumid| < \Delta\tau, \\
    0, & \text{otherwise,}
  \end{cases}
\end{equation}
where $\tau$ is the detector's proper time and $\taumid$ is the time at which the switching function reaches its peak value.
We use this switching function throughout this paper. Its Fourier transform is~\cite{ng_LittleExcitementHorizon_2022}
\begin{equation}\label{eq:hat-chi}
  \hat{\chi}(\omega)
  = \frac{\Delta\tau}{\pi} \, H\left(\omega \Delta\tau / \pi\right) e^{-i \omega \taumid},
\end{equation}
where
\begin{equation}\label{eq:H(z)}
  H(z) \defeq \frac{3 \sin(\pi z)}{z (1-z^2) (4 - z^2)}.
\end{equation}

\subsection{Stationary Trajectories}\label{sec:stationary}

A trajectory $\bx(\tau)$ is called \textit{stationary} with respect to the state of the field if the Wightman distribution $W(\tau, \tau')$ depends only on $\tau - \tau'$~\cite{letaw_StationaryWorldLines_1981}. Physically, this means the detector's response is invariant under time translations: only the shape and duration of the switching function matter, not where in the trajectory it is switched on. In this case, we write $W(\tau, \tau')$ as $W(\tau - \tau')$, and Eq.~\eqref{eq:F(E)-def} can be expressed as~\cite{fewster_WaitingUnruh_2016}
\begin{equation}\label{eq:F(E)-hatW}
    \response(E) = \frac{1}{2\pi} \int_{-\infty}^\infty d\omega \, |\hat{\chi}(\omega)|^2 \hat{W}(E + \omega),
\end{equation}
where $\hat{\chi}(\omega)$ is given in Eq.~\eqref{eq:hat-chi} and $\hat{W}(\omega)$ depends on the trajectory and state of the field.

In the long-duration limit ($\Delta\tau \to \infty$ at fixed $E$), Ref.~\cite{fewster_WaitingUnruh_2016} showed that
\begin{equation}\label{eq:F(E)-limit-adiabatic}
    \frac{\response(E)}{\Delta \tau}
    \to \frac{\hat{W}(E)}{2 \pi^2} \int_{-\infty}^\infty dz \, H^2(z)
    = \frac{35}{64} \hat{W}(E).
\end{equation}
In this limit, $\response(E)$ is proportional to the measurement duration, as expected for a detector in a stationary state. The time required for $\response$ to approach this limit can be understood as the detector's relaxation time, or the time it takes to ``thermalize''---that is, to reach equilibrium with the field~\cite{fewster_WaitingUnruh_2016}.
In the following section, we will study measurements made over durations shorter than the relaxation time, which will have important implications for detectors falling into black holes.

\section{Inertial Detectors in Minkowski Spacetime}\label{sec:minkowski}

Our primary goal is to study observers falling into black holes. However, such observers cannot make measurements over arbitrarily long durations---a detector dropped from any finite distance will hit the singularity in finite proper time. In fact, we will see that a particle detector switched on and off near the horizon can \textit{never} be in the long-duration limit discussed in Section~\ref{sec:stationary}. In physical terms, a detector falling into a black hole does not spend enough time in the near-horizon region to thermalize. Therefore, in order to interpret measurements made by infalling observers, we must first understand the behavior of particle detectors when the measurement duration is less than the relaxation time.

In this section we study inertial detectors in thermal states in Minkowski space. This setting has several advantages. First, Minkowski space has a well-defined ``particle'' concept shared by all inertial observers and the distribution of particles in a thermal state is well understood. In this familiar environment we can set aside, for now, the additional complications of black-hole spacetimes. Second, thermal states are homogeneous and isotropic, which makes inertial trajectories stationary, and Wightman distributions along inertial trajectories can be expressed as Fourier transforms of elementary functions. This reduces the task of calculating $\response$ to a one-dimensional integral, Eq.~\eqref{eq:F(E)-hatW}, which is significantly simpler and computationally cheaper than the black hole case. Finally, black hole vacuum states resemble thermal states~\cite{hawking_ParticleCreationBlack_1975,wald_ParticleCreationBlack_1975,hartle_PathintegralDerivationBlackhole_1976,unruh_NotesBlackholeEvaporation_1976}, at least asymptotically, so it is reasonable to conjecture that a freely falling observer, who sees no curvature locally, will perceive a state with similar characteristics to a Minkowski thermal state. Indeed, we will see in Section~\ref{sec:schwarz} that this holds true.

We consider a massless scalar field in 4-dimensional Minkowski spacetime, which can be expressed in the standard plane-wave basis~\cite{birrell_QuantumFieldsCurved_1982} as
\begin{equation}
  \phi(x) = \int \frac{d\vk}{\sqrt{16 \pi^3 \omega}} \left[a_\vk e^{-i k \cdot x} + a_\vk^\dagger e^{i k \cdot x}\right],
\end{equation}
where $x = (t, \vx)$, $k = (\omega, \vk)$, $\omega = |\vk|$, and $k \cdot x = \omega t - \vk \cdot \vx$, with the integral performed over all 3-dimensional space. Here, $a_\vk^\dagger$ and $a_\vk$ are the creation and annihilation operators, respectively, which satisfy the equal-time commutation relations
\begin{equation}
  \left[ a_\vk,  a_{\vk'}^\dagger \right] = \delta(\vk - \vk'), \quad \left[ a_\vk,  a_{\vk'} \right] = \left[ a_\vk^\dagger,  a_{\vk'}^\dagger \right] = 0.
\end{equation}
The \textit{Minkowski vacuum} state $\vacM$ is defined by $a_\vk \vacM = 0$ for all $\vk$. 
In a thermal state with temperature $T$, the positive-frequency Wightman distribution takes the form~\cite{birrell_QuantumFieldsCurved_1982,takagi_VacuumNoiseStress_1986}
\begin{equation}\label{eq:W-thermal-minkowski}
  W(x, x') = \int \frac{d\vk}{16 \pi^3 \omega} \left[ (1 + n_{\vk}) e^{-i k \cdot (x-x')} + n_{\vk} e^{i k \cdot (x-x')} \right],
\end{equation}
where
\begin{equation}
  n_{\vk} = \frac{1}{e^{\omega/T} - 1}.
\end{equation}

\subsection{Static Detector}

First we calculate the response of a particle detector at rest with respect to the isotropic thermal state. Substituting the trajectory $\bx(\tau) = (\tau, \vx_0)$ into Eq.~\eqref{eq:W-thermal-minkowski}, where $\vx_0$ is the location of the detector, the pullback of the Wightman distribution to the detector's worldline is $W(\tau,\tau') = W(\tau - \tau')$, where
\begin{align}
  W(s)
  &= \int_{-\infty}^\infty \frac{d\omega}{2\pi} \, e^{i \omega s} \, \frac{\omega}{2\pi (e^{\omega / T} - 1)},
\end{align}
which implies that
\begin{equation}\label{eq:What-thermal-stationary}
  \hat{W}(\omega) = \frac{\omega}{2 \pi (e^{\omega / T} - 1)}.
\end{equation}
Combining Eqs.~\eqref{eq:hat-chi}, \eqref{eq:F(E)-hatW}, and~\eqref{eq:What-thermal-stationary}, the response function is~\cite{ng_LittleExcitementHorizon_2022}
\begin{equation}\label{eq:F(E)-minkowski-static}
  \response
  = \frac{1}{4 \pi^2} \int_{-\infty}^{\infty} \frac{dz \, z}{1 - e^{-\pi z/(T \Delta \tau)}} H^2(z + E \Delta\tau / \pi),
\end{equation}
where $H$ is given in Eq.~\eqref{eq:H(z)}.

If we take $T \Delta\tau \to 0$ in Eq.~\eqref{eq:F(E)-minkowski-static}, we obtain~\cite{ng_LittleExcitementHorizon_2022}
\begin{equation}\label{eq:F-switch}
  \response \to
  \frac{1}{4 \pi^2} \int_{0}^{\infty} dz \, z \, H^2(z + E \Delta\tau / \pi).
\end{equation}
This limit can be viewed from two perspectives. First, it describes taking the temperature to zero while keeping the measurement duration constant. At zero temperature, the thermal state becomes the vacuum state, where no particles are present. Despite this, a detector in the vacuum state registers a nonzero response, even with a positive energy gap, due to the time-varying coupling between the detector and the field~\cite{birrell_QuantumFieldsCurved_1982,sriramkumar_ResponseFinitetimeParticle_1996}.
The $T \Delta\tau \to 0$ limit can also be viewed as making the measurement duration very short at a constant temperature. The response is asymptotically identical to a detector in the vacuum state, so it is independent of the number of particles in the field.
Therefore, in the $T \Delta\tau \to 0$ limit, the detector's response can be attributed solely to the effects of switching, with no contributions from the particles in the field.

Conversely, the long-duration limit (at fixed $E$ and $T$) is, by Eq.~\eqref{eq:F(E)-limit-adiabatic},
\begin{equation}\label{eq:F-par}
  \response \to
  \frac{35 \Delta\tau}{128 \pi} \frac{E}{e^{E / T} - 1},
\end{equation}
which is thermal in the KMS sense (Eq.~\eqref{eq:detailed-balance}) and proportional to the measurement duration, as expected for a particle detector in a thermal state.
In this limit, the effects of switching are negligible~\cite{birrell_QuantumFieldsCurved_1982,takagi_VacuumNoiseStress_1986}.

\subsubsection{Detecting Thermal Particles}

\begin{figure*}[t]
  \includegraphics{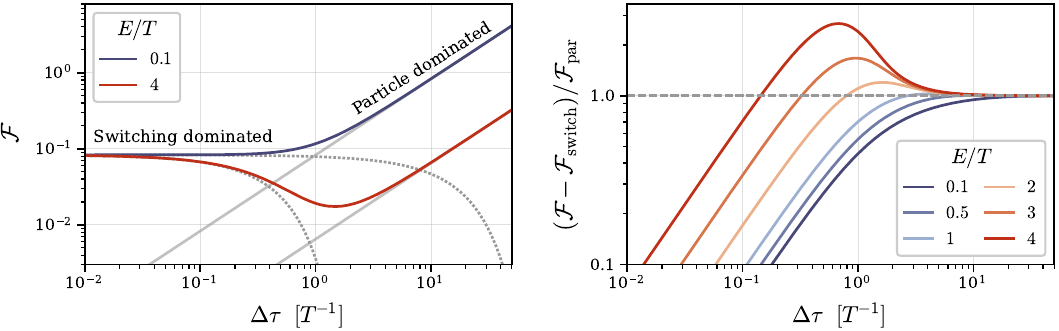}
  \caption{The response $\response$ of a static detector in a Minkowski thermal state with temperature $T$, as a function of the measurement duration $\Delta\tau$. \textbf{Left:} the transition between switching-dominated and particle-dominated regimes occurs at $\Delta\tau \sim T^{-1}$ when the detector energy gap $E$ is of order $T$. Gray lines denote the asymptotic limits for small (dotted) and large (solid) $\Delta\tau$.
  \textbf{Right:} $\response$ does \textit{not} simply decouple into its ``switching component'' $\response_\text{switch}$ plus its ``particle component'' $\response_\text{par}$. Depending on the detector energy and the switching duration, the actual response can be smaller or larger than this sum.
  }\label{fig:minkowski-F-vs-T-dtau-combined}
\end{figure*}

Consider using a particle detector to observe particles in a thermal Minkowski state. This is a simplified model for detecting Hawking radiation, which we will study in Section~\ref{sec:schwarz}. Observing particles involves detector excitations (rather than de-excitations), so we assume that $E>0$ throughout this section.

For short measurement durations, the detector's response is dominated by switching effects, with its asymptotic form given by Eq.~\eqref{eq:F-switch}. In this regime, we say that $\response$ is \textit{switching dominated} and denote it by $\response_\text{switch}$. Conversely, for long measurement durations, the response is dominated by the detection of thermal particles, with its asymptotic form given by Eq.~\eqref{eq:F-par}. In this regime, we say that $\response$ is \textit{particle dominated} and denote it by $\response_\text{par}$.

At what measurement duration does the response transition from switching dominated to particle dominated? Ref.~\cite{fewster_WaitingUnruh_2016} investigated a related question in the limit of large $E$, but our focus is on energies near the characteristic energy $T$ of the thermal state.

The left panel of Figure~\ref{fig:minkowski-F-vs-T-dtau-combined} shows $\response$ as a function of $\Delta\tau$ for two different values of $E$, obtained by numerically integrating Eq.~\eqref{eq:F(E)-minkowski-static}.
This plot is independent of $T$ (provided $T>0$) because Eq.~\eqref{eq:F(E)-minkowski-static} depends only on the dimensionless parameters $T \Delta\tau$ and $E \Delta\tau$.
For energies of order $T$, the transition between the switching-dominated and particle-dominated regimes occurs at $\Delta \tau \sim T^{-1}$.
Interestingly, this transition appears to happen even when $\Delta \tau < E^{-1}$, suggesting that the detector can become particle dominated even when the measurement duration is shorter than the period of the particles it is detecting.

The fact that a detector is switching dominated when $\Delta \tau \lesssim T^{-1}$ has significant implications for freely falling observers attempting to measure Hawking radiation. The timescale $\THawk^{-1}$, where $\THawk$ is the Hawking temperature of the black hole, is much longer than the time an infalling observer can spend near the horizon. Consequently, attempts by these observers to make localized, near-horizon measurements of Hawking radiation will generally be dominated by switching effects.

When switching effects dominate, it is unclear how to separate $\response$ into distinct ``switching'' and ``particle'' components. The right panel of Figure~\ref{fig:minkowski-F-vs-T-dtau-combined} shows that $\response$ does not simply separate into a sum of $\response_\text{switch}$ and $\response_\text{par}$.
For energies $E \lesssim T$, the total response is consistently \textit{less} than the sum of these terms across a wide range of $\Delta\tau$. Conversely, for energies $E \gtrsim T$, there is an intermediate range of $\Delta \tau$ where the response is actually \textit{greater} than this sum. These results suggest that, in the switching-dominated regime, it is not straightforward to decompose $\response$ into separate switching and particle components. For an observer falling into a black hole, this complicates attempts to observe Hawking radiation near the horizon, as switching effects dominate and it is unclear how to isolate the Hawking radiation signal from the overall response.

\subsubsection{Measuring the Temperature}

\begin{figure*}[t]
  \includegraphics{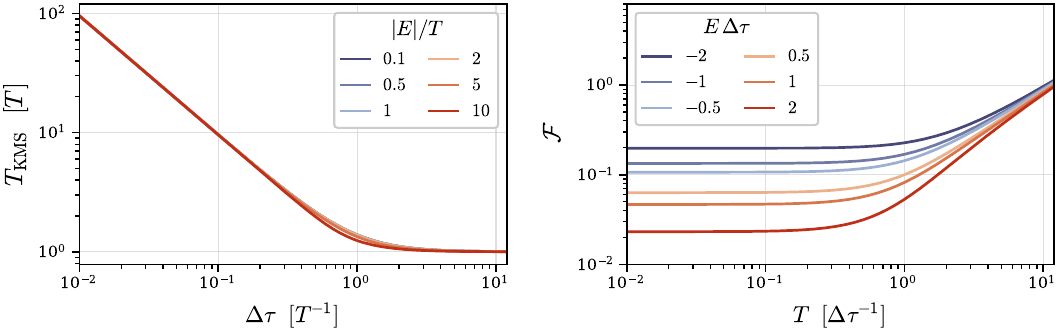}
  \caption{\textbf{Left:} the KMS temperature $\tKMS$ measured by a static detector is much larger than the temperature $T$ of the thermal state when $\Delta\tau \lesssim T^{-1}$. \textbf{Right:} the response $\response$ of a static detector is monotonic in $T$ when $E$ and $\Delta\tau$ are held fixed. Thus, $\response$ can be used to measure $T$, even for short switching durations where the detector has not yet thermalized.}\label{fig:minkowski-T_KMS-F}
\end{figure*}

Suppose that instead of detecting particles, we aim to measure the temperature {\red of the field}. We will later generalize this idea to black hole spacetimes.

One approach is to calculate the KMS temperature~\cite{juarez-aubry_OnsetDecay1+1_2014,hodgkinson_StaticDetectorsCirculargeodesic_2014,brenna_UnruhPhenomena_2016} by using two detectors to measure $\response(E)$ and $\response(-E)$. Rearranging Eq.~\eqref{eq:detailed-balance} and assuming the coefficient in Eq.~\eqref{eq:P(E)} is equal for $E$ and $-E$, we have
\begin{equation}\label{eq:T_KMS}
  \tKMS = \frac{E}{\ln (\response(-E) / \response(E))}.
\end{equation}
In the long-duration limit, Eq.~\eqref{eq:F-par} implies that $\tKMS \to T$. However, $\tKMS$ diverges in the short-duration limit because $\response(E) \to \response(-E)$ by Eq.~\eqref{eq:F(E)-minkowski-static}. {\red This behavior is illustrated in the left panel of Figure~\ref{fig:minkowski-T_KMS-F}, which shows that $\tKMS \gg T$ for $\Delta \tau \lesssim T^{-1}$ and $\tKMS \to T$ for $\Delta \tau \gtrsim T^{-1}$.  In other words, $\tKMS$ includes contributions from excitations due to switching, which dominate for $\Delta \tau \lesssim T^{-1}$.}
As an interesting aside, the detailed balance condition is approximately satisfied at this KMS temperature, but only for sufficiently small values of $E$. This suggests that, despite the divergence of $\tKMS$, a detector in the short-duration limit should not necessarily be considered out of equilibrium, at least if equilibrium is defined by the KMS condition~\cite{hodgkinson_StaticDetectorsCirculargeodesic_2014,brenna_UnruhPhenomena_2016,fewster_WaitingUnruh_2016}.
In this paper, we will consider the detector to have thermalized only when it satisfies Eq.~\eqref{eq:F(E)-limit-adiabatic}, in which case it must also satisfy the KMS condition.

{\red Alternatively, given the response $\response$ of a single detector with \textit{any} switching duration, it is possible to calculate the temperature $T$ (\textit{excluding} switching effects) by numerically inverting Eq.~\eqref{eq:F(E)-hatW}.} This is because $\response$ is monotonic in $T$ when $E$ and $\Delta \tau$ are held fixed, as illustrated in the right panel of Figure~\ref{fig:minkowski-T_KMS-F}. This allows us to {\red determine} $T$ even for short measurement durations where the detector has not yet thermalized. We will leverage this in Section~\ref{sec:schwarz} to define effective local temperatures along infalling trajectories in black-hole spacetimes.

\subsection{Constant-Velocity Detector}

We now extend the analysis of the previous section to detectors moving at constant velocities with respect to the thermal state. 
This is motivated by the idea that observers falling into black holes will measure Doppler-shifted Hawking radiation.
However, as discussed in the previous section, measurements by a particle detector falling into a black hole will be dominated by switching rather than Hawking radiation. How is a detector's response, when it is switching dominated, affected by its velocity relative to the field?

Consider a detector in 4-dimensional Minkowski space with trajectory $\bx = \vx_0 + \vv t$, where $t$ is the Minkowski time coordinate and $v \defeq |\vv| > 0$. The detector's proper time is $\tau = t / \gamma$, where $\gamma = (1 - v^2)^{-1/2}$ is the Lorentz factor. Substituting this trajectory into Eq.~\eqref{eq:W-thermal-minkowski} and performing the angular integrals in spherical coordinates, we obtain
\begin{equation}\label{eq:W-4D-minkowski-isotropic}
  W(s) = \int_{-\infty}^\infty \frac{d\omega}{2\pi} \, e^{i \gamma \omega s} \sinc(\gamma v \omega s) \, \frac{\omega}{2\pi(e^{\omega / T} - 1)},
\end{equation}
where $\sinc(x) \defeq \sin(x) / x$, with Fourier transform~\cite{costa_BackgroundThermalContributions_1995}
\begin{align}
  \hat{W}(\omega)
  &=  \frac{T}{4 \pi \gamma v} \log\left[ \frac{1 - e^{-(\omega/T) \sqrt{1 + v}/\sqrt{1 - v}}}{1 - e^{-(\omega/T) \sqrt{1 - v}/\sqrt{1 + v}}} \right]. \label{eq:W-hat-minkowski-inertial} 
\end{align}
The factors of $\sqrt{1 + v}/\sqrt{1 - v}$ and $\sqrt{1 - v}/\sqrt{1 + v}$ are Doppler blueshift and redshift factors, respectively, for waves traveling directly toward and away from the detector's trajectory~\cite{birrell_QuantumFieldsCurved_1982}.
In the limit as $v \to 0$, Eq.~\eqref{eq:W-hat-minkowski-inertial} reduces to Eq.~\eqref{eq:What-thermal-stationary}.

\begin{figure*}[t]
  \includegraphics{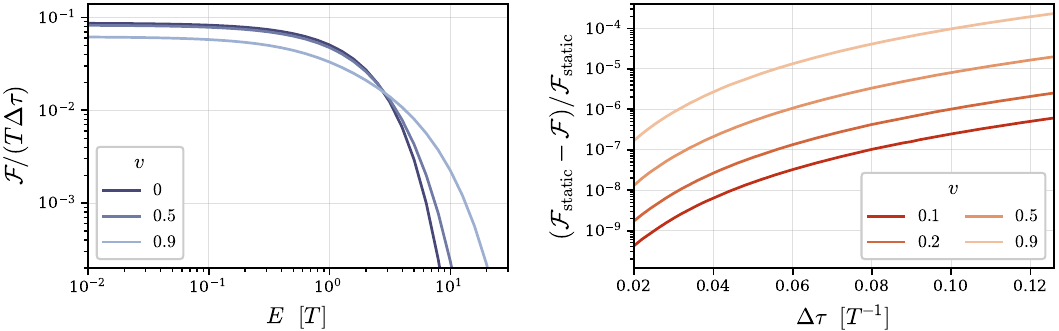}
  \caption{The response $\response$ of a detector moving with constant velocity $v$ through a Minkowski thermal state with temperature $T$, for very long measurement durations (left) and shorter durations (right). \textbf{Left:} in the limit of large $\Delta\tau$, the dependence of $\response$ on the energy gap $E$ is modified by the Doppler effect. \textbf{Right:} for small measurement durations, $\response$ is only weakly dependent on $v$ because switching effects dominate. $\response_\text{static}$ denotes the response measured by a static detector. The detector energy gap is $E = T^{-1}$, but the results are similar for all $E$ within an order of magnitude this value.}\label{fig:minkowski-inertial-plots}
\end{figure*}

The detector's response function is determined by Eqs.~\eqref{eq:F(E)-hatW} and~\eqref{eq:W-hat-minkowski-inertial}. In the long-duration limit, its velocity relative to the field induces a Doppler shift, as shown in the left panel of Figure~\ref{fig:minkowski-inertial-plots}. The Doppler shift causes $\response$ to be smaller than the response of a static detector for $E \ll T$ and larger for $E \gg T$ (see Ref.~\cite{hodgkinson_StaticDetectorsCirculargeodesic_2014} for a relevant discussion).
However, we are more interested in shorter measurement durations where switching effects are significant, as for a detector falling into a black hole.

The right panel of Figure~\ref{fig:minkowski-inertial-plots} shows the detector's response for measurement durations comparable to those needed for localized measurements by an infalling observer near the horizon of a black hole. Since switching effects dominate, $\response$ is only weakly dependent on the detector's velocity. For example, a detector traveling at 90\% the speed of light with $\Delta\tau < 0.1 / T$ will observe a relative change in $\response$ by a factor of only \num{e-4} compared to a static detector. Even this small change does not seem to be interpretable as a Doppler shift, because $\response$ decreases with increasing $v$ regardless of whether $E$ is larger or smaller than $T$, whereas the Doppler-shifted steady-state spectrum increases with increasing $v$ when $E \gg T$. This further highlights the difficulty of interpreting any part of $\response$ as measuring real particles when the detector is in the switching-dominated regime.

\section{Detector Falling into a Schwarzschild Black Hole}\label{sec:schwarz}

We now turn to our main goal: studying Hawking radiation from the perspective of an observer freely falling into a black hole. We begin by reviewing the Schwarzschild spacetime, massless scalar fields, and the standard Schwarzschild vacuum states. This discussion is not new, but it establishes notation and key concepts for our calculations to follow. We then derive the response of a freely falling Unruh-DeWitt detector and describe our numerical implementation for computing it. Readers primarily interested in the results may proceed directly to Section~\ref{sec:results}.

\subsection{Schwarzschild Spacetime}

The Schwarzschild metric is
\begin{equation}\label{eq:ds2-schwarz}
  ds^2 =  \left(1 - 2M / r \right) dt^2 - \frac{1}{1-2M / r} dr^2 - r^2 d\Omega^2,
\end{equation}
where $r > 2M$ and $d\Omega^2 \defeq d\theta^2 + \sin^2\theta \, d\varphi^2$ is the metric of the unit 2-sphere. It is the vacuum solution to the Einstein field equation outside of a spherically symmetric matter distribution of total mass $M$.

The Regge-Wheeler tortoise coordinate $r_*$ is defined by the relation
\begin{equation}\label{eq:drstar-dr}
  \dd{r_*}{r} = \frac{1}{1 - 2M/r}.
\end{equation}
The outgoing and ingoing null coordinates are defined by
\begin{equation}
  u = r - r_*, \quad v = t + r_*,
\end{equation}
and are so named because radial outgoing/ingoing null geodesics are lines of constant $u$/$v$. Finally, the Kruskal-Szekeres coordinates are defined by
\begin{equation}
  U = -e^{-u/(4M)}, \quad V = e^{v / (4M)}.
\end{equation}
In the Kruskal-Szekeres coordinate system $(U,V,\theta,\varphi)$, the Schwarzschild metric takes the form
\begin{equation}\label{eq:ds2-kruskal}
    ds^2 = \frac{32 M^3}{r} \, e^{-r/(2M)} dU \, dV - r^2 d\Omega^2,
\end{equation}
where $r$ is the solution to
\begin{equation}\label{eq:UV-r}
    UV = \left(1 - \frac{r}{2M}\right) e^{r/2M}.
\end{equation}
In this coordinate system, the Schwarzschild spacetime corresponds to $U < 0$ and $V > 0$. By extending the ranges of $U$ and $V$ to all real values for which $UV < 1$, we obtain the \textit{maxim extended Schwarzschild} (or \textit{Kruskal-Szekeres}) spacetime. We will henceforth refer to the maximally extended spacetime as the Schwarzschild spacetime. Its conformal diagram can be found in Refs.~\cite{birrell_QuantumFieldsCurved_1982,misner_Gravitation_2017}. The spacetime consists of four regions:
\begin{enumerate}[(i)]
  \item Region I (${U<0}$, ${V>0}$) is the original Schwarzschild spacetime;
  \item Region II (${U>0}$, ${V>0}$, ${UV < 1}$) is the black hole interior;
  \item Region III (${U>0}$, ${V<0}$) is identical to Region I, but the two regions are causally disconnected;
  \item Region IV (${U<0}$, ${V<0}$, ${UV < 1}$) is the white hole interior.
\end{enumerate}

The Schwarzschild spacetime has four independent Killing vectors. Three of these are spacelike wherever they are nonvanishing, and the remaining one is
\begin{equation}
  K = -\frac{U}{4M} \, \del_U + \frac{V}{4M} \, \del_V,
\end{equation}
which reduces to $\del_t$ in Region I. It is timelike in Regions I and III and spacelike in Regions II and IV.

\subsection{Radial Geodesics}

Freely falling particle detectors follow timelike geodesics. We are interested in trajectories that start in Region I (outside the black hole) and end up in Region II (inside the black hole), for which it is convenient to use ingoing Eddington-Finkelstein coordinates $(v, r, \theta, \varphi)$. The Schwarzschild metric in this coordinate system is
\begin{equation}
  ds^2 = \left(1 - 2M/r\right) dv^2 - 2 dv \, dr - r^2 d\Omega^2.
\end{equation}

The detector's trajectory is $\bx(\tau)$, where $\tau$ is its proper time. Since we only consider radial trajectories, we can set $\theta=\varphi=0$ without loss of generality. Then the 4-velocity $\dot\bx$ satisfies
\begin{equation}\label{eq:radial-geodesic-x2}
  1 = \dot{\bx}^2 = (1 - 2M / r) \dot{v}^2 - 2 \dot{v} \dot{r} ,
\end{equation}
where dots indicate derivatives with respect to $\tau$. Since $K = \del_v$ is a Killing vector, the following quantity is conserved along geodesics,
\begin{equation}\label{eq:radial-geodesic-E}
  \mathcal{E} \defeq K \cdot \dot\bx = (1 - 2M/r) \dot{v} - \dot{r}.
\end{equation}
Rearranging Eqs.~\eqref{eq:radial-geodesic-x2} and~\eqref{eq:radial-geodesic-E} gives the following system of equations governing a general radial timelike geodesic,
\begin{subequations}\label{eq:radial-geodesic-eqns}
  \begin{align}
    \dd{r}{\tau} &= \eta \sqrt{2M/r - 2M/R}, \\[5pt]
    \dd{v}{\tau} &= \frac{\sqrt{1 - 2 M / R} + \eta \sqrt{2M / r - 2M / R}}{1 - 2M / r},
  \end{align}
\end{subequations}
where $\eta = -1$ for an ingoing trajectory and $+1$ for an outgoing trajectory, and where we have defined $R \defeq 2M / (1 - \mathcal{E}^2)$. If $R>0$, it is the Schwarzschild radius at which the detector is instantaneously at rest. The limit $R \to \infty$ describes the case where the detector is at rest at infinity (i.e., it has zero mechanical energy), in which case Eq.~\eqref{eq:radial-geodesic-eqns} has a simple closed-form solution~\cite{misner_Gravitation_2017}. For finite $R$, there is a parametric solution~\cite{misner_Gravitation_2017}, but in this case we find it easier to solve Eq.~\eqref{eq:radial-geodesic-eqns} numerically.

\subsection{Classical Scalar Field}

A massless scalar field $\phi(x)$ satisfies the Klein-Gordon equation,
\begin{equation}\label{eq:klein-gordon} 
  \nabla_\mu \nabla^\mu \phi = 0.
\end{equation}
We now review its general solution in the Schwarzschild spacetime.

\subsubsection{Region I}

\begin{figure}[t]
  \includegraphics{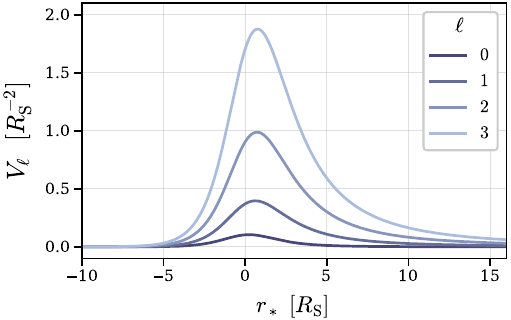}
  \caption{The effective potential $V_\ell$ for $\ell \leq 4$.}\label{fig:V_ell}
\end{figure}

To begin, we focus on Region I only. In Schwarzschild coordinates $(t, r, \theta, \varphi)$, the Klein-Gordon equation is separable and its general solution is a linear combination of the basis modes~\cite{dewitt_QuantumFieldTheory_1975}
\begin{equation}\label{eq:kg-modes}
  \Phi_{\omega \ell}(r) Y_{\ell m} (\theta, \varphi) e^{\pm i \omega t}
\end{equation}
for $\omega > 0$, $\ell \in \{ 0, 1, ... \}$, $m \in \{ -\ell, ..., \ell \}$, where $Y_{\ell m}$ is a spherical harmonic~\cite{NIST_spherical_harmonics}, and where $\Phi_{\omega \ell}$ is a solution to
\begin{equation}\label{eq:radialeq-R-r}
  \dd[2]{\Phi}{r} + \frac{2 \left(r - M \right)}{r \left(r - 2M \right)} \dd{\Phi}{r}
  + \left[\frac{\omega^2 r^2}{\left(r - 2M \right)^2} - \frac{\ell(\ell + 1)}{r \left(r - 2M \right)}\right]\Phi = 0,
\end{equation}
which is known as the generalized spheroidal wave equation~\cite{leaver_SolutionsGeneralizedSpheroidal_1986a}. The basis modes in Eq.~\eqref{eq:kg-modes} proportional to $e^{- i \omega t}$ are positive frequency with respect to the timelike Killing vector $K = \del_t$, whereas those proportional to $e^{+i \omega t}$ are negative frequency.

By defining $\rho(r) \defeq r \, \Phi(r)$, Eq.~\eqref{eq:radialeq-R-r} takes the form of the time-independent Schr\"odinger equation in terms of the tortoise coordinate $r_*$,
\begin{equation}\label{eq:radialeq-u-rstar}
  \dd[2]{\rho}{r_*} = \left[V_\ell(r) - \omega^2 \right] \rho,
\end{equation}
with the effective potential
\begin{equation}\label{eq:V(r)}
    V_\ell(r) \defeq \left(1-\frac{2M}{r}\right)\left( \frac{\ell(\ell + 1)}{r^2} + \frac{2M}{r^3}\right).
\end{equation}
Figure~\ref{fig:V_ell} shows $V_\ell$ as a function of $r_*$ for $\ell \leq 4$. As we will discuss later, a key challenge in this paper is solving Eq.~\eqref{eq:radialeq-u-rstar} numerically across a large number of $(\omega, \ell)$ pairs.

Since $V_\ell(r)$ vanishes asymptotically close to the horizon ($r \to 2M$) and far from the black hole ($r \to \infty$), in those asymptotic regions, Eq.~\eqref{eq:radialeq-u-rstar} implies that $\rho_{\omega\ell}(r)$ is a linear combination of $e^{\pm i \omega r_*}$. We choose two independent solutions defined by their asymptotic behavior
\begin{equation}\label{eq:rho-in}
  \rho^\text{in}_{\omega\ell}(r) \to
  \begin{cases}
      B^\text{in}_{\omega \ell} \, e^{-i \omega r_*}, & r \to 2M, \\
      e^{-i \omega r_*} + A^\text{in}_{\omega \ell}  \, e^{+i \omega r_*}, & r \to \infty,
  \end{cases}
\end{equation}
and 
\begin{equation}\label{eq:rho-up}
  \rho^\text{up}_{\omega\ell}(r) \to
  \begin{cases}
      A^\text{up}_{\omega \ell}   \, e^{-i \omega r_*}+ e^{+i \omega r_*}, & r \to 2M, \\
      B^\text{up}_{\omega \ell}  \, e^{+i \omega r_*}, & r \to \infty.
  \end{cases}
\end{equation}
The $A$ and $B$ terms are reflection and transmission coefficients if Eq.~\eqref{eq:radialeq-u-rstar} is viewed as a one-dimensional scattering problem. Their values are implied by the coefficients of the other terms. These coefficients satisfy a number of identities, as summarized in Ref.~\cite{dewitt_QuantumFieldTheory_1975}.

We then obtain two independent solutions to Eq.~\eqref{eq:radialeq-R-r},
\begin{equation}\label{eq:Phi-inup}
  \Phi_{\omega\ell}^\text{in/up}(r) \defeq {\rho_{\omega\ell}^\text{in/up}(r)}/{r},
\end{equation}
and define the normalized positive-frequency modes
\begin{align}\label{eq:u-inup-I}
  u^\text{in/up}_{\omega \ell m}(t,r,\theta,\varphi) &\defeq \frac{Y_{\ell m}(\theta,\varphi)}{\sqrt{4 \pi \omega}} \Phi^\text{in/up}_{\omega \ell}(r) e^{-i \omega t}.
\end{align}
In Region I, the set $\{ u_{\omega\ell m}^\text{in/up}, u_{\omega\ell m}^{\text{in/up}*} \}$ for all $\omega,\ell,m$ is a complete set of solutions to the Klein-Gordon equation.
These modes satisfy the orthonormality relations
\begin{subequations}\label{eq:u-orthonormality}
  \begin{align}
    (u^j_{\omega\ell m}, u^{j'}_{\omega' \ell' m'}) &= \delta_{jj'} \delta_{\ell \ell'} \delta_{m m'} \delta(\omega - \omega'), \\
    (u^{j*}_{\omega\ell m}, u^{j'*}_{\omega' \ell' m'}) &= - \delta_{jj'} \delta_{\ell \ell'} \delta_{m m'} \delta(\omega - \omega'), \\
    (u^j_{\omega\ell m}, u^{j'*}_{\omega' \ell' m'}) &= 0,
  \end{align}
\end{subequations}
where $j,j' \in \{ \text{in}, \text{up} \}$, with respect to the Klein-Gordon inner product~\cite{birrell_QuantumFieldsCurved_1982}
\begin{equation}\label{eq:kg-inner-prod}
  (\phi_1, \phi_2) \defeq -i \int_\Sigma d\Sigma \, \sqrt{|g_\Sigma|} \, n^\mu \left[\phi_1 \nabla_\mu \phi_2^* - \phi_2^* \nabla_\mu \phi_1\right],
\end{equation}
where $\Sigma$ is a spacelike hypersurface with induced metric $g_\Sigma$ and $n$ is a future-directed unit vector orthogonal to $\Sigma$. In Region I, we can choose $\Sigma$ to be the $t=0$ hypersurface and $n \propto \del_t$. 

The modes $u^\text{in}_{\omega \ell m}$ are called \textit{in modes} because near the horizon they have the asymptotic behavior
\begin{equation}
  u^\text{in}_{\omega \ell m} \propto e^{-i \omega (t + r_*)}, \quad r \to 2M,
\end{equation}
which represent waves traveling into the horizon with no outgoing component. These waves originate infinitely far from the black hole at early times, traveling toward the horizon. They scatter off the gravitational potential of the black hole, with part of each wave traveling into the horizon and part traveling back out to infinity. Meanwhile, the modes $u^\text{up}_{\omega \ell m}$ are called \textit{up modes} because far from the black hole they have the asymptotic behavior
\begin{equation}
  u^\text{up}_{\omega \ell m} \propto e^{-i \omega (t - r_*)}, \quad r \to \infty,
\end{equation}
which represent waves traveling away from the black hole with no ingoing component. These waves originate on the past horizon of the white hole, traveling away from the horizon. Upon scattering off the potential, part of each wave travels back toward the horizon and part travels out to infinity. A helpful visualization of these modes can be found in Ref.~\cite{dewitt_QuantumFieldTheory_1975}.

\subsubsection{Regions II--IV}

In each of Regions II--IV, we introduce local Schwarzschild coordinates~\cite{fulling_AlternativeVacuumStates_1977} covering only that region, starting with
\begin{equation}
  t \defeq 2M \ln\left|V / U\right| , \quad r_* \defeq 2M \ln\left|UV\right|,
\end{equation}
where $t \in (-\infty, \infty)$ in each region, while $r_* \in (-\infty, 0)$ in Regions II and IV and $r_* \in (-\infty, \infty)$ in Region III. We define the local $r$ coordinate by Eq.~\eqref{eq:UV-r}, which means that $r \in (0, 2M)$ in Regions II and IV and $r \in (2M, \infty)$ in Region III. The local $r$ and $r_*$ coordinates satisfy Eq.~\eqref{eq:drstar-dr}.

In each region, in terms of these local coordinates, the metric takes the same form as Eq.~\eqref{eq:ds2-schwarz}. Therefore, the solution to the Klein-Gordon equation is again determined by Eqs.~(\ref{eq:kg-modes}--\ref{eq:V(r)}).

In Region III, all the equations take the same form as in Region I, so the general solution has the same form as in the previous section. However, in this region, the Killing vector $\del_t$ is \textit{past}-directed with respect to the global time-orientation of the extended Schwarzschild spacetime. Accordingly, we define positive frequency modes with respect to the future-directed Killing vector $\del_{-t} = -\del_t$ by taking the complex conjugate of Eq.~\eqref{eq:u-inup-I}, defining
\begin{align}\label{eq:v-inup-I}
  v^\text{in/up}_{\omega \ell m}(t,r,\theta,\varphi) &\defeq \frac{Y_{\ell m}^*(\theta,\varphi)}{\sqrt{4 \pi \omega}} \Phi^{\text{in/up}*}_{\omega \ell}(r) e^{i \omega t}.
\end{align}
Then $\{ v_{\omega\ell m}^\text{in/up}, v_{\omega\ell m}^{\text{in/up}*} \}$ is a complete set of solutions to the Klein-Gordon equation and satisfies the orthonormality relations in Eq.~\eqref{eq:u-orthonormality} with $u$ replaced with $v$.

In Regions II and IV, the potential $V_\ell$ in Eq.~\eqref{eq:V(r)} diverges at the singularity ($r = 0$) and vanishes asymptotically close to the horizon ($r \to 2M$). In the latter limit, as in the previous section, $\rho_{\omega\ell}(r)$ is a linear combination of $e^{\pm i \omega r_*}$, so for consistency with Eq.~\eqref{eq:rho-in}, we define the following solution by its asymptotic behavior near the horizon,
\begin{equation}
  \rho^\text{in}_{\omega\ell}(r) \to B^\text{in}_{\omega \ell} \, e^{-i \omega r_*}, \quad r \to 2M,
\end{equation}
where $B^\text{in}_{\omega \ell}$ is the transmission coefficient from Region I. We define $\Phi_{\omega\ell}^\text{in}$ in the same way as Region I. There is no analog of $\Phi^\text{up}_{\omega\ell}$ in Regions II and IV. Instead, we choose the second independent solution to Eq.~\eqref{eq:radialeq-R-r} to be
$\Phi^{\text{in}*}_{\omega\ell}$.

We have now solved the Klein-Gordon equation separately in all quadrants of the extended Schwarzschild spacetime. In the next section, we will quantize the scalar field by combining these per-quadrant solutions into different basis sets covering the entire spacetime.

\subsection{Quantum Scalar Field}

To quantize a classical scalar field theory, we take a complete set of orthonormal basis modes $\{ u_\beta(x), u_\beta^*(x) \}$, where $\beta$ represents the set of labels for each mode, and expand the field as~\cite{birrell_QuantumFieldsCurved_1982}
\begin{equation}
  \phi(x) = \sum_\beta \left[ b_\beta u_\beta(x) + b_\beta^\dagger u_\beta^*(x) \right],
\end{equation}
where $b_\beta^\dagger$ and $b_\beta$ are respectively the creation and annihilation operators for particles in mode $\beta$, which satisfy the commutation relations
\begin{equation}
  [b_\beta, b_{\beta'}] = 0, \quad [b_\beta^\dagger, b_{\beta'}^\dagger] = 0, \quad [b_\beta, b_{\beta'}^\dagger] = \delta_{\beta\beta'}.
\end{equation}
The vacuum state $\ket{0}$ is defined by $b_\beta \ket{0} = 0$ for all $\beta$.

Unlike Minkowski space where the global timelike Killing vector selects a preferred basis of plane waves, in a general curved spacetime there is no preferred basis and therefore no preferred vacuum state. Different choices of basis may correspond to different vacuum states that do not agree on the particle content of the field. In this section, we review the 3 standard vacuum states for the Schwarzschild spacetime. For more comprehensive reviews of these vacuum states, see Refs.~\cite{fulling_AlternativeVacuumStates_1977,candelas_VacuumPolarizationSchwarzschild_1980,birrell_QuantumFieldsCurved_1982}.

\subsubsection{Boulware Vacuum}

The Boulware vacuum~\cite{boulware_QuantumFieldTheory_1975} is obtained by quantizing the field using basis modes that are positive-frequency with respect to the Killing vectors $\del_t$ in Region I and $\del_{-t}$ in Region III, aligning with the natural definition of positive frequency for observers far from the black hole.

Let $u^\text{in/up}$ and $v^\text{in/up}$ denote the positive-frequency basis modes in Regions I and III, respectively, as defined in Eqs.~\eqref{eq:u-inup-I} and~\eqref{eq:v-inup-I}. We extend the domain of these modes to the union of Regions I and III, as follows
\begin{align}\label{eq:u-inup-I-III}
  u^\text{in/up}_{\omega \ell m} \defeq
  \begin{cases}
      u^\text{in/up}_{\omega \ell m}, &\text{Region I},\\
      0, &\text{Region III},\\
  \end{cases}
\end{align}
and
\begin{align}\label{eq:v-inup-I-III}
  v^\text{in/up}_{\omega \ell m} \defeq
  \begin{cases}
      0, &\text{Region I},\\
      v^\text{in/up}_{\omega \ell m}, &\text{Region III.}\\
  \end{cases}
\end{align}

Each of these functions can be analytically extended to a solution of the Klein-Gordon equation covering the entire extended Schwarzschild spacetime~\cite{boulware_QuantumFieldTheory_1975}.
In Kruskal-Szekeres coordinates in Regions I and II, the in modes have the form~\cite{hodgkinson_ParticleDetectorsCurved_2013}
\begin{equation}\label{uv-in-KS}
  \begin{aligned}
    u^\text{in}_{\omega \ell m} &=
    \frac{Y_{\ell m}(\theta,\varphi) }{\sqrt{4 \pi \omega}}
    \tilde\Phi^\text{in}_{\omega \ell}(r) V^{-i 4M \omega}, &r > 0,\\
    v^\text{in}_{\omega \ell m} &= 0, &r > 0,
  \end{aligned}
\end{equation}
and the up modes have the form
\begin{align}
  u^\text{up}_{\omega \ell m} &=
  \frac{Y_{\ell m}(\theta,\varphi)}{\sqrt{4 \pi \omega}} \times
  \begin{cases}
    \tilde\Phi^\text{up}_{\omega \ell}(r) (-U)^{i 4M \omega}, &r > 2M,\\[10pt]
    \dfrac{A^\text{up}_{\omega \ell}}{B^\text{in}_{\omega \ell}} \tilde\Phi^\text{in}_{\omega \ell}(r) V^{-i 4M \omega}, &r < 2M, \nonumber\\
  \end{cases}\\[6pt]
  v^\text{up}_{\omega \ell m} &=
  \frac{Y_{\ell m}^*(\theta,\varphi)}{\sqrt{4 \pi \omega}} \times
  \begin{cases}
    0, & r > 2M,\\[6pt]
    \dfrac{1}{B^{\text{in}}_{\omega \ell} } \tilde\Phi^{\text{in}}_{\omega \ell}(r) U^{-i 4M \omega}, &r < 2M, \label{eq:uv-up-KS}
  \end{cases}
\end{align}
where
\begin{align}\label{eq:tildePhi-inup}
  \tilde\Phi^\text{in}_{\omega\ell}(r) \defeq e^{+i \omega r_*} \Phi^\text{in}_{\omega\ell}(r), \quad
  \tilde\Phi^\text{up}_{\omega\ell}(r) \defeq e^{-i \omega r_*} \Phi^\text{up}_{\omega\ell}(r). 
\end{align}

These modes and their complex conjugates are a complete orthonormal set of solutions to the Klein-Gordon equation across the entire spacetime. Quantizing the field with respect to these modes, we have
\begin{equation}
  \begin{aligned}
    \phi(x) = \sum_{\ell, m} \int_{0}^\infty d\omega \, \left[ b^\text{in}_{\omega \ell m} u^\text{in}_{\omega \ell m} + b^\text{up}_{\omega \ell m} u^\text{up}_{\omega \ell m} \right. \\
    \left. + b^{'\text{in}}_{\omega \ell m} v^\text{in}_{\omega \ell m} + b^{'\text{up}}_{\omega \ell m} v^\text{up}_{\omega \ell m} + \text{h.c.} \right],
  \end{aligned}
\end{equation}
where ``h.c.'' represents the Hermitian conjugates of all preceding terms. The Boulware vacuum $\ket{0_\text{B}}$ is defined by $b^\text{in/up}_{\omega \ell m} \ket{0_\text{B}} = 0$ and $b^{'\text{in/up}}_{\omega \ell m} \ket{0_\text{B}} = 0$.
\vspace{3pt}

The Boulware vacuum reduces to the Minkowski vacuum at large distances from the black hole~\cite{boulware_QuantumFieldTheory_1975}. However, it is not physically realistic near the horizon; for example, its stress-energy tensor diverges as $r \to 2M$ in a freely falling frame~\cite{christensen_TraceAnomaliesHawking_1977,candelas_VacuumPolarizationSchwarzschild_1980}. Since our main goal is to study observers falling through the horizon, we will focus instead on vacuum states that give more physically realistic descriptions near the horizon.

\subsubsection{Hartle-Hawking Vacuum}

The Hartle-Hawking vacuum~\cite{hartle_PathintegralDerivationBlackhole_1976,israel_ThermofieldDynamicsBlack_1976} is the unique vacuum state that is regular everywhere and invariant under the Schwarzschild Killing vector $\del_t$. It is defined by the normalized modes~\cite{hodgkinson_ParticleDetectorsCurved_2013,hodgkinson_StaticDetectorsCirculargeodesic_2014}
\begin{equation}
  w^\text{in/up}_{\omega \ell m} \defeq \frac{u^\text{in/up}_{\omega \ell m} + e^{-4 \pi M \omega} v^{\text{in/up}*}_{\omega \ell m}}{\sqrt{1 - e^{-8 \pi M \omega}}}
\end{equation}
and
\begin{equation}
  \bar{w}^\text{in/up}_{\omega \ell m} \defeq \frac{e^{- 4\pi M \omega} u^{\text{in/up}*}_{\omega \ell m} + v^{\text{in/up}}_{\omega \ell m}}{\sqrt{1 - e^{-8 \pi M \omega}}},
\end{equation}
which, along with their complex conjugates, form a complete orthonormal set of solutions to the Klein-Gordon equation. Quantizing the field with respect to these modes, we have
\begin{equation}
  \begin{aligned}
    \phi(x) = \sum_{\ell, m} \int_{0}^\infty d\omega \, \left[ d^\text{in}_{\omega \ell m} w^\text{in}_{\omega \ell m} + d^\text{up}_{\omega \ell m} w^\text{up}_{\omega \ell m} \right. \\
    \left. + \bar{d}^{\text{in}}_{\omega \ell m} \bar{w}^\text{in}_{\omega \ell m} + \bar{d}^{\text{up}}_{\omega \ell m} \bar{w}^\text{up}_{\omega \ell m} + \text{h.c.} \right],
  \end{aligned}
\end{equation}
with the Hartle-Hawking vacuum $\ket{0_\text{H}}$ defined by $d^\text{in/up}_{\omega \ell m} \ket{0_\text{H}} = 0$ and $\bar{d}^{\text{in/up}}_{\omega \ell m} \ket{0_\text{H}} = 0$.
\vspace{3pt}

Distant observers perceive the Hartle-Hawking vacuum as a bath of thermal radiation at the Hawking temperature~\cite{hartle_PathintegralDerivationBlackhole_1976}. This vacuum state represents a black hole in thermal equilibrium with its surroundings, such as when enclosed within a sufficiently small reflecting cavity~\cite{birrell_QuantumFieldsCurved_1982}.

\subsubsection{Unruh Vacuum}

The Unruh vacuum~\cite{unruh_NotesBlackholeEvaporation_1976} is designed to reproduce the state of an astrophysical black hole formed from gravitational collapse. The field is quantized as~\cite{hodgkinson_ParticleDetectorsCurved_2013,hodgkinson_StaticDetectorsCirculargeodesic_2014}
\begin{equation}
  \begin{aligned}
    \phi(x) = \sum_{\ell, m} \int_{0}^\infty d\omega \, \left[ b^\text{in}_{\omega \ell m} u^\text{in}_{\omega \ell m} + d^\text{up}_{\omega \ell m} w^\text{up}_{\omega \ell m} \right. \\
    \left. + b^{'\text{in}}_{\omega \ell m} v^\text{in}_{\omega \ell m} + \bar{d}^{\text{up}}_{\omega \ell m} \bar{w}^\text{up}_{\omega \ell m} + \text{h.c.} \right],
  \end{aligned}
\end{equation}
with the Unruh vacuum $\ket{0_\text{U}}$ defined by $b^\text{in}_{\omega \ell m} \ket{0_\text{U}}$ = 0, $d^\text{up}_{\omega \ell m} \ket{0_\text{U}} = 0$, $b^{'\text{in}}_{\omega \ell m} \ket{0_\text{U}} = 0$, and $\bar{d}^{\text{up}}_{\omega \ell m} \ket{0_\text{U}} = 0$.
\vspace{3pt}

The Unruh vacuum represents a flux of thermal particles at the Hawking temperature emanating from the black hole, with no ingoing flux from infinity~\cite{birrell_QuantumFieldsCurved_1982}.
It emerges in models of gravitational collapse in the late-time, near-horizon limit~\cite{fabbri_ModelingBlackHole_2005}, and is thus a physically realistic model for observers falling into astrophysical black holes sufficiently long after their formation.

In the following sections, we will study measurements by freely falling observers in both the Hartle-Hawking and Unruh vacuum states, which are regular across the future horizon. Although the Unruh vacuum is the most accurate description of the quantum state of astrophysical black holes, the Hartle-Hawking vacuum will enable a simple operational definition of the local temperature along the trajectory of an infalling observer, yielding insights that may apply more broadly to black holes formed from gravitational collapse.

\subsection{Detector Response on a Radial Trajectory}

We are now ready to compute the response of a particle detector falling radially into a Schwarzschild black hole.
The expressions in this section were previously derived by Ref.~\cite{ng_LittleExcitementHorizon_2022}. We review them here for completeness with minor notational differences for conciseness and numerical stability.

Given a vacuum state defined by a set of orthonormal basis modes $\{ u_\beta(x), u_\beta^*(x) \}$, the positive frequency Wightman distribution can be expressed as
\begin{equation}\label{eq:W(x,x')-modesum}
    W(x,x') \defeq \braket{0|\phi(x) \phi(x') | 0} = \sum_\beta u_\beta(x) u_{\beta}^*(x'),
\end{equation}
where $\ket{0}$ is the vacuum state associated with that basis. The detector response along the trajectory $\bx(\tau)$ is obtained by substituting Eq.~\eqref{eq:W(x,x')-modesum} into Eq.~\eqref{eq:F(E)-def} to get
\begin{equation}\label{eq:F(omega)-modesum}
    \response = \sum_{\beta} \left| \int_{-\infty}^\infty d\tau \, \chi(\tau) e^{-iE \tau} u_\beta(\bx(\tau)) \right|^2.
\end{equation}

It is now simply a matter of plugging in the basis modes $\{ u_\beta(x) \}$ for each vacuum of interest. On a radial trajectory, we can assume without loss of generality that $\theta = \varphi = 0$, allowing us to simplify the modes using the identity~\cite{NIST_spherical_harmonics}
\begin{equation}\label{eq:Ylm-theta0}
    Y_{\ell m}(\theta,\varphi) = 
    \begin{cases}
        \sqrt{\dfrac{2 \ell + 1}{4 \pi}}, & m=0, \\
        0, & m>0.
    \end{cases}
\end{equation}

\begin{widetext}
  The detector responses in the Boulware, Hartle-Hawking, and Unruh states are respectively given by 
  \begin{equation}
    \responseB = \FB^\text{in} + \FB^\text{up}, \quad
    \responseH = \FH^\text{in} + \FH^{\overline{\text{in}}} + \FH^\text{up} + \FH^{\overline{\text{up}}}, \quad
    \responseU = \FB^\text{in} + \FH^\text{up} + \FH^{\overline{\text{up}}},
\end{equation}
where
\begin{subequations}\label{eq:F-B-H-in-up}
\begin{align}
    \FB^\text{in/up}
    &= \sum_{\ell = 0}^\infty \int_0^\infty d\omega \frac{2 l + 1}{16 \pi^2 \omega} \left| \int_{-\infty}^\infty d\tau \, \chi(\tau) e^{-i E \tau} I^\text{in/up}_{\omega\ell}(\tau) \right|^2, \\[5pt]
    \FH^{\text{in/up},\overline{\text{in/up}}}
    &= \sum_{\ell = 0}^\infty \int_0^\infty \frac{d\omega}{1 - e^{-8 \pi M \omega}} \frac{2 l + 1}{16 \pi^2 \omega} \left| \int_{-\infty}^\infty  d\tau \, \chi(\tau) e^{-i E \tau} I^{\text{in/up},\overline{\text{in/up}}}_{\omega\ell}(\tau) \right|^2, 
\end{align}
\end{subequations}
and
\begingroup
\allowdisplaybreaks
\begin{subequations}\label{eq:I-in-up}
\begin{align}
    I_{\omega\ell}^\text{in}(\tau) &= \tilde\Phi_{\omega\ell}^\text{in}(r) V^{-i4M\omega}, \\[5pt]
    I_{\omega\ell}^{\overline{\text{in}}}(\tau) &= e^{-4 \pi M \omega} \tilde\Phi_{\omega\ell}^{\text{in}*}(r) V^{i 4 M \omega}, \\[5pt]
    I_{\omega\ell}^\text{up}(\tau)
    &=
    \begin{cases}
        \tilde\Phi_{\omega\ell}^\text{up}(r) (-U)^{i4M\omega}, & r > 2M, \\[5pt]
        \displaystyle
        \frac{A_{\omega\ell}^\text{up}}{B_{\omega\ell}^\text{in}} \tilde\Phi_{\omega\ell}^\text{in}(r)V^{-i 4 M \omega} + 
        e^{-4 \pi M \omega}\frac{1}{B_{\omega\ell}^{\text{in}*}} \tilde\Phi_{\omega\ell}^{\text{in}*}(r) U^{i 4 M \omega}, & r < 2M,
    \end{cases} \\[5pt]
    I_{\omega\ell}^{\overline{\text{up}}}(\tau)
    &=
    \begin{cases}
      e^{-4 \pi M \omega} \tilde\Phi_{\omega\ell}^{\text{up}*}(r) (-U)^{-i 4 M \omega}, & r > 2M, \\[5pt]
      \displaystyle
      e^{-4 \pi M \omega} \frac{A_{\omega\ell}^{\text{up}*}}{B_{\omega\ell}^{\text{in}*}} \tilde\Phi_{\omega\ell}^{\text{in}*}(r) V^{i 4 M \omega} + 
      \frac{1}{B_{\omega\ell}^\text{in}} \tilde\Phi_{\omega\ell}^\text{in}(r) U^{-i 4 M \omega}, & r < 2M.
  \end{cases}
\end{align}
\end{subequations}
\endgroup
where $U,V$ and $r$ are evaluated on the detector trajectory $\bx(\tau)$.

\vspace{5em}
\end{widetext}

\subsection{Numerical Implementation}\label{sec:numerical}

We now briefly summarize our procedure for numerically computing the response of a particle detector on a radial trajectory near the horizon of a Schwarzschild black hole.
We use a modified version of the methods described in Refs.~\cite{hodgkinson_ParticleDetectorsCurved_2013,ng_LittleExcitementHorizon_2022}. Full details are available in our publicly accessible code~\cite{shallue_HawkingRadiationHorizon_2024}. In the code, and consequently throughout this section, we use units in which $c=\hbar=k_B=\RS=1$.

First, we solve for the detector's trajectory $\bx(\tau)$.
All trajectories we study are timelike radial geodesics labeled by $\eta$ and $R$, where $\eta = -1$ for ingoing trajectories and $+1$ for outgoing trajectories, and $R > 2M$ is the radial Schwarzschild coordinate at which the detector is instantaneously at rest. For given values of $\eta$ and $R$, we obtain $\bx(\tau)$ by numerically solving the system of differential equations in Eq.~\eqref{eq:radial-geodesic-eqns}.

For a given trajectory, we typically want to compute the detector's response at multiple locations, which requires specifying a sequence of switching functions. Each switching function $\chi(\tau)$ is defined by Eq.~\eqref{eq:chi-cos} and is parametrized by $\taumid$ and $\Delta \tau$. We choose a fixed value of $\Delta \tau$ for all measurements and specify a sequence of radial coordinates $\rmid$ where the switching functions will be centered. We then numerically solve for the $\taumid$ values that satisfy $r(\taumid) = \rmid$ along the trajectory $\bx(\tau)$. 
These $\taumid$ values determine the sequence of switching functions to be evaluated.

Given the trajectory $\bx(\tau)$, switching functions $\chi(\tau)$, and a detector energy gap $E$, we now turn to the integral expressions in Eq.~\eqref{eq:F-B-H-in-up}.
To evaluate these, we must contend with two infinite parameter ranges. 
We handle the sum over angular modes $\ell$ by truncating it at a maximum value $\ell_\text{max}$. To ensure convergence, we require that the summand at $\ell = \ell_\text{max}$ is at least a factor of \num{e8} smaller than the largest summand for $\ell < \ell_\text{max}$.
For the infinite integral over  $\omega$ for each $\ell$, we truncate the integration at a maximum value $\omega_\text{max}$ and compute the integral using the trapezoidal rule on a grid spanning $(0, \omega_\text{max})$. The sampling density is manually adjusted as a function of $\ell$ to ensure sufficient resolution in regions where the integrand varies rapidly. We set $\omega_\text{max}$ dynamically, stopping when the result changes by less than $\epsilon$ over the preceding interval of width $\Delta \omega = 10 / \RS$, with $\epsilon$ chosen as a function of $\ell$.
We have verified that our results do not change significantly if we increase the $\omega$ sampling or the values of $\ell_\text{max}$ and $\omega_\text{max}$.

For each $(\ell, \omega)$ pair, we must compute an integral over the detector's proper time $\tau$. Although this integral has infinite terminals, it only needs to be evaluated over the finite interval during which the detector is active (i.e., $|\tau - \taumid| < \Delta\tau$). We use the \texttt{QUADPACK}~\cite{piessens_QuadpackSubroutinePackage_1983} routine for automatic numerical integration, as implemented in \texttt{scipy}~\cite{virtanen_SciPy10Fundamental_2020}, to perform this calculation. We set the absolute and relative error tolerances to \num{e-14} and \num{e-8}, respectively, with a maximum of \num{e3} subintervals in the adaptive algorithm. These values balance accuracy with computational efficiency. We confirmed that adjusting these parameters does not significantly affect our results.

The most numerically challenging part of these calculations is computing the functions $\tilde\Phi^\text{in/up}_{\omega\ell}$ and the reflection and transmission coefficients $A^\text{up}_{\omega\ell}$ and $B^\text{in}_{\omega\ell}$.
The functions $\tilde\Phi^\text{in/up}_{\omega\ell}$ are obtained by numerically solving Eq.~\eqref{eq:radialeq-u-rstar} with appropriate boundary conditions and applying Eqs.~\eqref{eq:Phi-inup} and~\eqref{eq:tildePhi-inup}. To compute the boundary conditions, we use power series expansions as $r \to 2M$ for the in modes and $r \to \infty$ for the up modes, as described in Ref.~\cite{hodgkinson_ParticleDetectorsCurved_2013}. We truncate the series at 101 and 51 terms for the in and up modes, respectively. In Ref.~\cite{hodgkinson_ParticleDetectorsCurved_2013}, Eq.~\eqref{eq:radialeq-u-rstar} was solved numerically for $\num{e-7} \lesssim r/ \RS \lesssim \num{e4}$, but the power series expansions remain accurate over most of this interval, so we solve Eq.~\eqref{eq:radialeq-u-rstar} numerically over a much smaller range of $r$, resulting in a significant computational speedup. We use an explicit Runge-Kutta method of order 8~\cite{hairer_SolvingOrdinaryDifferential_2008}, as implemented in \texttt{scipy}, which yields a 7th degree interpolation polynomial accurate to 7th order. We set the absolute and relative error tolerances to 0 and \num{e-10}, respectively, to balance accuracy with computational efficiency.

In Region I, numerical issues can arise when $\omega^2$ is much smaller than the peak value of $V_\ell$ in Eq.~\eqref{eq:radialeq-u-rstar}, which is the case for a significant portion of the required $(\ell, \omega)$ values. In this scenario, the potential barrier is much larger than the energy of the incident waves, so the amplitude of the transmitted wave is much smaller than the reflected wave---potentially by tens of orders of magnitude. In our implementation, we take care to identify and avoid issues arising from numerical precision, particularly when performing arithmetic operations with numbers of very different magnitudes.

Additional numerical challenges arise very close to the horizon. The functions in Eq.~\eqref{eq:I-in-up} oscillate infinitely rapidly in the detector's proper time as it approaches the horizon (and immediately after it crosses the horizon). We therefore exclude a small region around the horizon from our integration. The excluded region is $|\tau - \tau_\text{horiz}| < \delta$, where $\tau_\text{horiz}$ is the proper time at which the detector crosses the horizon. We use $\delta = \num{e-6} \RS$, and we have verified that our results do not change if we make $\delta$ as small as $\num{e-10} \RS$. Decreasing $\delta$ significantly increases the runtime of the numerical integrator, as it must resolve increasingly high frequency oscillations. We validated our integration across the horizon in two additional ways. First, we changed the integration variable from $\tau$ to $r_*$ to mitigate the issue of increasing oscillation frequency. Second, we changed our integration algorithm to an 8th order Runge-Kutta method, which integrates along the trajectory instead of recursively dividing it into subintervals. In all cases, our results remained consistent.

At each location on the trajectory, we compute all of the terms in Eq.~\eqref{eq:F-B-H-in-up} to obtain the detector response $\response$ in all three vacuum states. Our method requires approximately \num{e5} different $(\omega, \ell)$ pairs to ensure convergence of the $\ell$-sums and $\omega$-integrals. Accordingly, for each trajectory, we need to solve Eq.~\eqref{eq:radialeq-u-rstar} approximately \num{e5} times. Then for a given detector energy gap $E$, we need to perform up to four integrals over $\tau$ for each $(\omega,\ell)$ pair and measurement location $\taumid$. Fewer integrals are required if some of the outer $\omega$-integrals have already converged. We typically evaluate around twenty locations along each trajectory, requiring a total of approximately \num{e7} $\tau$-integrals along the entire trajectory, per energy gap $E$. The overall processing time is dominated by numerically solving Eq.~\eqref{eq:radialeq-u-rstar} and performing the integrals over $\tau$. As a point of reference, the results shown in Figure~\ref{fig:schwarz-F-vacua} in the next section took around nine hours of wall-clock time per energy gap $E$ running continuously on twenty parallel processes across two Intel{\textregistered} Xeon{\textregistered} E5-2630v4 2.2 GHz CPUs.

\section{Results}\label{sec:results}

\subsection{Response of a Freely Falling Detector}

\begin{figure*}[t]
  \includegraphics{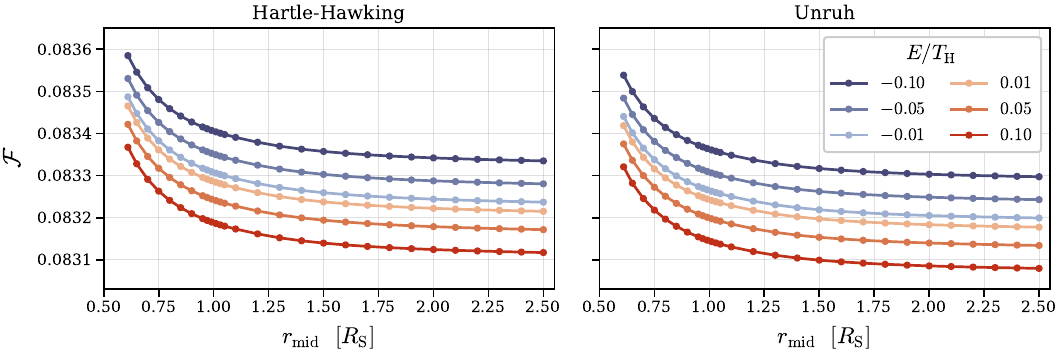}
  \caption{The response $\response$ of a detector freely falling from rest at infinity into a Schwarzschild black hole in the Hartle-Hawking and Unruh vacuum states. At all energies shown, the response increases smoothly as the detector approaches and crosses the horizon. The legend is the same for both panels.}\label{fig:schwarz-F-vacua}
\end{figure*}

We first consider a detector freely falling into a Schwarzschild black hole on a radial trajectory from rest at infinity (i.e., with $R = \infty$).
Figure~\ref{fig:schwarz-F-vacua} shows the response in the Hartle-Hawking and Unruh vacuum states as a function of the switching midpoint $\rmid$, as calculated using the methods described in the previous section. Each measurement is integrated over a proper-time duration of $2 \Delta \tau = 0.05 / \THawk$, which ensures that each measurement is localized near the horizon.
We evaluated a range of positive and negative energy gaps $E$ with magnitudes near the Hawking temperature, which is the characteristic energy of Hawking radiation. The results are qualitatively similar for each $E$, with the magnitude of $\response$ decreasing as $E$ increases. The responses in the two vacuum states are similar, with the Unruh vacuum response slightly lower in magnitude than the Hartle-Hawking vacuum response.

For all energy gaps considered, $\response$ increases smoothly as the detector approaches and crosses the horizon, with no distinctive features at the horizon. This contrasts with the findings of Ref.~\cite{ng_LittleExcitementHorizon_2022}, who observed a ``bump'' in $\response$ near the horizon. In Appendix~\ref{appendix:extra-plots} (Figure~\ref{fig:schwarz-Ng-Fig1}), we directly compare our results to Ref.~\cite{ng_LittleExcitementHorizon_2022}, finding agreement away from the horizon but without the near-horizon bump.

How should the measurements in Figure~\ref{fig:schwarz-F-vacua} be interpreted? Do they represent Hawking radiation, or are they mainly due to switching the detector on and off?
Recall from Section~\ref{sec:minkowski} that a detector in a Minkowski thermal state is switching dominated when $\Delta\tau \lesssim T^{-1}$. Here we have $\Delta \tau = 0.025 / \THawk$, suggesting that the detector is likely to be switching dominated, at least if its local state is approximately thermal near the Hawking temperature.
In the Hartle-Hawking vacuum, this is the case for large $r$ because the state is asymptotically thermal at the Hawking temperature~\cite{hartle_PathintegralDerivationBlackhole_1976,birrell_QuantumFieldsCurved_1982}.
{\red Near the horizon, as we will argue in the next section, the Hartle-Hawking vacuum has an effective local temperature approximately twice the Hawking temperature}, suggesting that the response remains switching dominated as the detector approaches the horizon.
Moreover, we will show that the response near the horizon depends only weakly on $\Delta \tau$, rather than being proportional to $\Delta \tau$ as would be expected if it were particle dominated.
We conclude that, throughout the trajectory, the detector's response {\red in Figure~\ref{fig:schwarz-F-vacua}} primarily reflects a local interaction with the field due to switching and does not provide a direct measurement of Hawking radiation.

Our results imply that a detector freely falling from rest at infinity \textit{cannot} make localized measurements of Hawking radiation near the horizon without being dominated by switching effects. This is because choosing $\Delta \tau \gg \THawk^{-1}$ results in measurements integrated over an interval much larger than $10 \RS$. Later in this section, we will consider detectors falling from rest at finite distances, allowing larger $\Delta \tau$ while still keeping the measurements localized. However, even in those cases, the detector will remain switching dominated: it is not possible for an infalling observer to measure Hawking radiation near the horizon without interference from switching, at least when the energy gap is on the order of the Hawking temperature.
Despite this, in the next section, we will show that the detector's response is still useful for measuring the effective local temperature along the trajectory.

\subsection{Measuring the Effective Local Temperature}

\begin{figure*}[t]
  \includegraphics{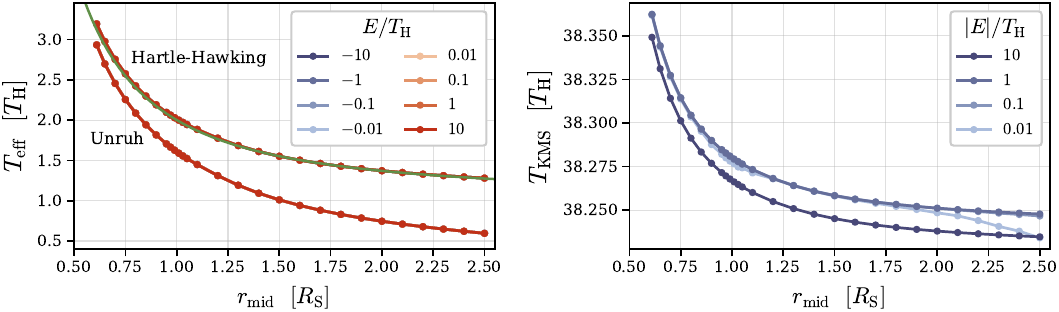}
  \caption{Two definitions of local temperature measured by a detector freely falling from rest at infinity into a Schwarzschild black hole, using a switching duration of $\Delta\tau = 0.025/\THawk$. 
  \textbf{Left:} the effective temperature $\teff$ in the Hartle-Hawking and Unruh states, which is consistent for all detector energy gaps $E$ in this plot. In the Hartle-Hawking vacuum, it is well-fit by Eq.~\eqref{eq:T_brynjolfsson}, shown by the green line.
  \textbf{Right:} {\red the KMS temperature in the Hartle-Hawking vacuum is much larger than the effective temperature because it includes contributions from switching.
  The small deviation in the curve for $|E|=0.01 \, \THawk$ is due to numerical challenges at small $|E|$ and large $\rmid$.}
  }\label{fig:schwarz-T-KMS-eff}
\end{figure*}

We propose a straightforward operational definition of effective temperature along a trajectory, which is motivated by our study of detectors in thermal Minkowski states in Section~\ref{sec:minkowski}.
First, we measure the response $\response$ of an Unruh-DeWitt detector with positive or negative energy gap $E$ along the trajectory, using a switching duration $\Delta\tau$ that is sufficiently short to ensure a localized measurement.
For a freely falling detector near the horizon of a Schwarzschild black hole, this requires choosing a short enough $\Delta\tau$ that $\response$ will dominated by switching effects.
However, we can identify the following well-defined quantity:

\begin{description}
  \item[Effective temperature, $\teff$] The  Minkowski thermal temperature $T$ at which a static detector, using the same values of $E$ and $\Delta\tau$, would record the same response at a specified location on the trajectory.
\end{description}
For fixed values of $E$ and $\Delta\tau$, $\teff$ is unique because the response of a static detector in a Minkowski thermal state is a monotonic function of temperature, even when switching effects dominate (see Figure~\ref{fig:minkowski-T_KMS-F}). While in principle $\teff$ depends on $E$ and $\Delta\tau$, we will see that the dependence is sufficiently weak that it can be neglected.

While this definition can, in principle, be used to assign an effective temperature to any trajectory in any quantum state, it is most meaningful in states expected to locally resemble isotropic thermal states, such as the Hartle-Hawking vacuum.
This is because approximating the detector's environment by a Minkowski thermal state may be inappropriate for highly anisotropic or nonthermal states. For example, the Unruh vacuum represents only outgoing Hawking radiation, so this definition of effective temperature may not accurately describe the experience of an observer.
Moreover, the effective temperature can only be determined if the measured response is greater than the Minkowski vacuum response, but this is not guaranteed in an arbitrary state.
For example, near the horizon, an infalling detector in the Boulware vacuum measures a smaller response than a static detector in the Minkowski vacuum~\cite{ng_LittleExcitementHorizon_2022}. The Boulware vacuum is not physically realistic near the horizon~\cite{christensen_TraceAnomaliesHawking_1977,candelas_VacuumPolarizationSchwarzschild_1980} and cannot be assigned an effective temperature using this method.

\begin{figure*}[t]
  \includegraphics{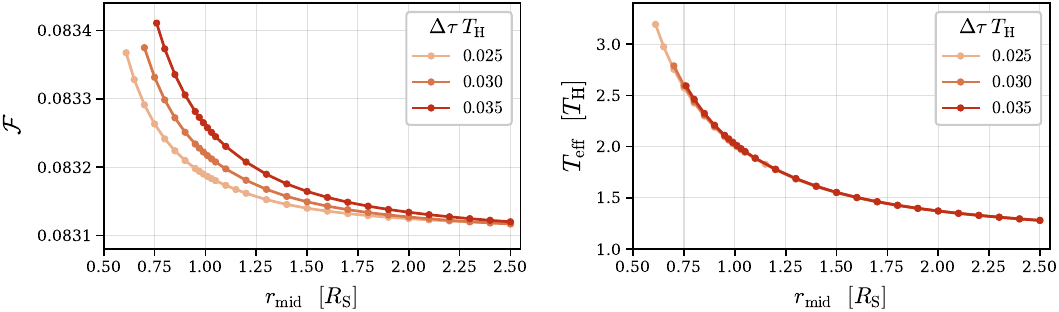}
  \caption{For a detector freely falling into a Schwarzschild black hole in the Hartle-Hawking state, changing the switching duration $\Delta\tau$ changes the measured response $\response$ (left panel), but the effective temperature $\teff$ (right panel) remains consistent along the trajectory. Here the detector energy gap is $E = 0.1 \, \THawk$. The results are similar for other energy gaps.}\label{fig:schwarz-F-and-T-vs-r-vary-dtau}
\end{figure*}

The left panel of Figure~\ref{fig:schwarz-T-KMS-eff} shows the effective temperature measured by an infalling detector in the Hartle-Hawking and Unruh vacuum states. Since the detector is not in true equilibrium, the effective temperature is not guaranteed to be consistent for all values of the energy gap $E$. Nonetheless, at all measurement locations considered, the effective temperature is consistent to within one part in \num{e3} for a range of positive and negative energy gaps spanning three orders of magnitude around the Hawking temperature.
The effective temperature is also robust to changes in $\Delta\tau$, provided the duration remains small enough for measurements to be reasonably localized. As shown in Figure~\ref{fig:schwarz-F-and-T-vs-r-vary-dtau}, the effective temperature is consistent for $\Delta\tau = 0.025 / \THawk$, $0.03 / \THawk$, and $0.035 / \THawk$, despite the dependence of $\response$ on $\Delta \tau$. As discussed in the previous section, the fact that $\response$ is not proportional to $\Delta \tau$ in Figure~\ref{fig:schwarz-F-and-T-vs-r-vary-dtau} indicates that the detector is switching dominated.

Far from the black hole, the effective temperature measured by an infalling detector in the Hartle-Hawking vacuum approaches the Hawking temperature, as would be expected. As the detector approaches the horizon, the effective temperature rises smoothly, reaching approximately twice the Hawking temperature at the horizon. This increase continues smoothly as the detector crosses into the black hole's interior.
In the Unruh vacuum, the trend is similar, but the effective temperature is lower because $\response$ is smaller than in the Hartle-Hawking vacuum. However, as noted earlier, our definition of effective temperature may not be appropriate for the Unruh vacuum, which is not an isotropic state. Extending our definition of effective temperature to anisotropic states would be an interesting direction for future work.

Remarkably, the effective temperature measured by an infalling detector in the Hartle-Hawking vacuum closely matches Eq.~\eqref{eq:T_brynjolfsson}, which was derived by embedding 4-dimensional Schwarzschild spacetime into 6-dimensional Minkowski spacetime~\cite{brynjolfsson_TakingTemperatureBlack_2008}.
Both approaches relate the local state along the trajectory to an ``equivalent'' Minkowski state.
However, our approach is based on the precise response of a detector in 4-dimensional spacetimes, accounting for the detector's energy gap and the profile of its switching function.
In contrast, Eq.~\eqref{eq:T_brynjolfsson} is derived by associating a temperature with the higher-dimensional Unruh effect, even though the detector's trajectory lacks the uniform acceleration required for the Unruh effect and measurements by detectors in the two spaces do not match exactly~\cite{brynjolfsson_TakingTemperatureBlack_2008,langlois_ImprintsSpacetimeTopology_2005}.
The agreement between these distinct approaches might be coincidental, but it may also indicate that global embedding methods hold promise for gaining deeper insights into the near-horizon state of black holes.

For comparison with our effective temperature, the right panel of Figure~\ref{fig:schwarz-T-KMS-eff} shows the KMS temperature, defined by Eq.~\eqref{eq:T_KMS}, as measured along the trajectory. {\red As discussed in Section~\ref{sec:minkowski}, the KMS temperature includes contributions from excitations due to switching, so it is much higher than the effective temperature.}
{\red Since $\tKMS$ is dominated by switching effects, it depends quite strongly on $\Delta\tau$. For example, changing $\Delta\tau$ from $0.025/\THawk$ to $0.035 /\THawk$ changes $\tKMS$ from around $38 \,\THawk$ to $27 \,\THawk$, because the switching effects diminish. Additionally, the slight energy dependence of $\tKMS$ shows that the detector's response is not strictly thermal in the KMS sense.  Similar to the effective temperature, the KMS temperature increases smoothly as the detector approaches and crosses the horizon.}

{\red Why does the local temperature {\red ($\teff$ or $\tKMS$)} measured by a freely falling  detector increase as it approaches the horizon?}

%
%
{\red It is well known that a \textit{static} detector in the Hartle-Hawking vacuum measures a temperature that increases as the Schwarzschild radius decreases, with the temperature diverging at the horizon~\cite{unruh_NotesBlackholeEvaporation_1976,candelas_VacuumPolarizationSchwarzschild_1980,hodgkinson_StaticDetectorsCirculargeodesic_2014}. This is due to the Unruh effect~\cite{unruh_NotesBlackholeEvaporation_1976}, as a detector at a fixed Schwarzschild radius has a constant proper acceleration that increases and becomes infinite as the detector approaches the horizon. However, freely falling detectors are not accelerating, so they do not undergo excitations due to acceleration.}

{\red The temperature of a classical, static fluid in a gravitational field has a so-called Tolman gradient as measured by Killing observers~\cite{tolman_WeightHeatThermal_1930,tolman_TemperatureEquilibriumStatic_1930,santiago_TolmanlikeTemperatureGradients_2018}. Recent studies have attempted to predict the local temperature of Hawking radiation by incorporating quantum corrections into the Tolman gradient~\cite{gim_QuantalTolmanTemperature_2015,eune_EffectiveTolmanTemperature_2017,kim_EffectiveTolmanTemperature_2017}, but these predictions do not match our results.
}


{\red It is sometimes suggested that Hawking radiation has a temperature gradient due to the Doppler effect~\cite[e.g.][]{barbado_HawkingRadiationPerceived_2011,barbado_HawkingRadiationPerceived_2012,smerlak_NewPerspectivesHawking_2013}.
When an infalling detector approaches the horizon, its velocity relative to static observers increases, approaching the speed of light at the horizon. The argument goes that outgoing Hawking radiation will be blueshifted in the detector's frame, increasing the measured temperature.
On the other hand, we have shown that localized measurements by freely falling observers are dominated by switching effects, and that switching effects are only weakly sensitive to velocity.
To investigate this hypothesis, we recalculated the temperatures shown in the left panel of Figure~\ref{fig:schwarz-T-KMS-eff}, adjusting for the {\red detector's velocity}.} Specifically, we determined the Minkowski thermal temperature at which an inertial detector moving at the same speed as the infalling detector (relative to a static observer) would measure the same value of $\response$. We found that the effective temperature is almost the same regardless of whether we adjust for {\red velocity} or not, except very close to the horizon where the velocity approaches the speed of light.
%

Ultimately, we do not yet have a satisfactory explanation for the gradient in the response function and temperature measured by an infalling detector. The increasing response \textit{inside} the horizon is particularly intriguing, especially in the Unruh vacuum, which is not supposed to contain any ingoing radiation modes. Understanding this may require a more abstract conceptualization of the near-horizon state than as a flux of Hawking radiation particles, particularly given the ambiguity in defining particles near the horizon~\cite{wald_ParticleCreationBlack_1975,davies_EnergymomentumTensorEvaporating_1976,christensen_TraceAnomaliesHawking_1977}.

\subsection{Detectors on Other Radial Trajectories}

We conclude our numerical analysis with a brief investigation of detectors on other radial trajectories near the horizon of a Schwarzschild black hole.

First, we consider releasing a detector from rest near the horizon, rather than from infinity. This detector will fall slower than one dropped from infinity, meaning that for a given measurement duration $\Delta\tau$ (as measured in the detector's frame), it will cover a shorter distance. This allows the detector to make more localized measurements for a given value of $\Delta\tau$.

\begin{figure*}[t]
  \includegraphics{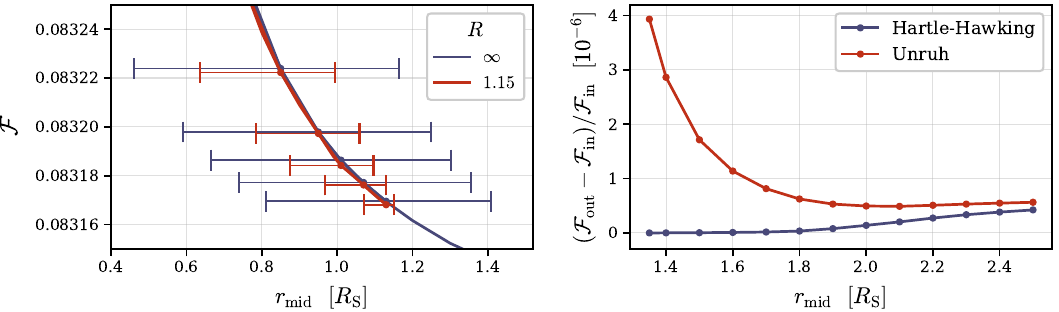}
  \caption{Measurements by detectors on other radial trajectories near the horizon of a Schwarzschild black hole. Each detector has an energy gap of $E = 0.1 \, \THawk$ and duration of $\Delta\tau = 0.025 / \THawk$, as measured in their local reference frames. The field is in the Hartle-Hawking vacuum in the left panel and either the Hartle-Hawking or Unruh vacuum in the right panel.
  \textbf{Left:} a detector dropped from rest at $R = 1.15 \, \RS$ falls more slowly than a detector dropped from infinity and therefore makes more localized measurements. Horizontal lines indicate the regions over which the measurements are made. \textbf{Right:} the response $\response_\text{out}$ measured by a detector on an outgoing trajectory compared to the response $\response_\text{in}$ measured on an infalling trajectory.}\label{fig:schwarz-F-vs-r-other-radial}
\end{figure*}

The left panel of Figure~\ref{fig:schwarz-F-vs-r-other-radial} compares the response of a detector dropped from rest at $R = 1.15 \, \RS$ in the Hartle-Hawking vacuum with that of a detector dropped from rest at infinity. Both detectors have a switching duration of $\Delta\tau = 0.025 / \THawk$. The horizontal lines around each point indicate the region over which the measurements are made. Since the detector dropped closer to the horizon moves more slowly, its measurements are significantly more localized. Despite this, the measured response is very similar between the two trajectories, suggesting that a detector freely falling from infinity provides a sufficiently local measure of the region near the horizon.

For a detector dropped near the horizon, one might wonder whether $\Delta\tau$ can be made large enough for the detector to become particle dominated while still making localized measurements, thereby enabling direct detection of Hawking radiation. Unfortunately, this is not the case.
For example, to obtain a measurement centered at $\rmid = 1.1 \RS$ and localized to a region of width $\RS$ in $r$, the maximum possible value of $\Delta\tau$ on an infalling trajectory is approximately $0.1 / \THawk$, corresponding to a detector dropped from rest at $R \approx 1.3 \RS$. However, we saw in Section~\ref{sec:minkowski} that a detector needs $\Delta\tau \gg \THawk^{-1}$ to become particle dominated. We conclude that a freely falling Unruh-DeWitt detector cannot directly measure Hawking radiation near the horizon of a Schwarzschild black hole, at least for energies on the order of the Hawking temperature.

Finally, we consider detectors on radially \textit{outgoing} trajectories, which can be achieved by launching them outward from near the horizon. In the Hartle-Hawking vacuum, we expect an outgoing detector to measure the same response as an infalling detector with the same value of $R$ due to the time-reversal invariance of this state. However, the Unruh vacuum represents a time-asymmetric flux emanating from the black hole, so we expect the response of an outgoing detector to differ from that of an infalling detector.

The right panel of Figure~\ref{fig:schwarz-F-vs-r-other-radial} compares the responses of ingoing and outgoing detectors with $R = \infty$ in both the Hartle-Hawking and Unruh vacua. Each detector has an energy gap of $E = 0.1 \, \THawk$ and a measurement duration of $\Delta\tau = 0.025 / \THawk$, as measured in their local reference frames.
In the Hartle-Hawking vacuum, the responses along the two trajectories differ by only a few parts in \num{e7}, which is likely a numerical artifact caused by differences in the convergence rates of the mode sums.
In the Unruh vacuum, however, the outgoing detector measures a larger response than the ingoing detector by a few parts in \num{e6}, with this difference becoming more pronounced near the horizon.
Although these detectors are switching dominated and do not directly measure Hawking radiation, this behavior is \textit{opposite} to what we would expect from a Doppler shift. For outgoing Hawking radiation, the frequency of a given wave would be lower in the outgoing detector's frame and higher in the ingoing detector's frame. With both detectors using the same value of $E$, the outgoing detector should measure higher-frequency Hawking radiation, which is less abundant than lower-frequency radiation, suggesting its response should be smaller---yet the opposite is true.
This further highlights the challenge of interpreting particle-detector measurements near the horizon in terms of Hawking radiation.

\section{Conclusions}\label{sec:conclusions}

We conducted a detailed analysis of what freely falling observers would measure near the horizon of a semiclassical Schwarzschild black hole. One of our key findings is that an Unruh-DeWitt detector cannot directly measure Hawking radiation near the horizon. This is due to the fact that the time required for the detector to ``thermalize''---that is, for its response to become dominated by the detection of actual particles rather than by switching effects---is much longer than the time it spends near the horizon. While this conclusion assumes that the local state experienced by the detector is well approximated by a Minkowski thermal state, we found it consistently supported by detailed calculations of the detector's response in the vicinity of the horizon.

By numerically evaluating the response of a detector along an infalling trajectory, we demonstrated that the response increases smoothly as it approaches the horizon and continues rising within the black hole's interior. This trend holds for a broad range of positive and negative detector energy gaps centered around the Hawking temperature. Our findings differ from those of Ref.~\cite{ng_LittleExcitementHorizon_2022}, who observed a localized ``bump'' in the detector's response near the horizon. However, our results align with theirs both outside and inside the black hole, away from the immediate vicinity of the horizon.

We proposed a straightforward, operational definition for the effective temperature along a trajectory, applicable to states that locally resemble thermal states, such as the Hartle-Hawking vacuum.
We found that the effective local temperature measured by a freely falling detector in the Hartle-Hawking vacuum increases smoothly from the Hawking temperature far from the horizon to twice the Hawking temperature at the horizon, and continues to rise into the black hole's interior.
Contrary to common suggestions in the literature, we found that the increase in effective temperature near the horizon is not due to Doppler shifting of Hawking radiation.

Intriguingly, our effective temperature, computed from exact calculations of detector measurements with specific energy gaps and switching profiles, closely aligns with the local temperature given in Eq.~\eqref{eq:T_brynjolfsson}. This equation was obtained by embedding Schwarzschild spacetime into a higher-dimensional Minkowski space and calculating an approximate Unruh temperature along the trajectory---an approach not originally expected to precisely describe the experience of an infalling observer~\cite{brynjolfsson_TakingTemperatureBlack_2008}.
The close agreement between our numerical results and this higher-dimensional embedding method suggests that further exploration of such embeddings could provide additional insights into the near-horizon behavior of black holes.

\vspace*{1.2em}

\section*{acknowledgments}
  The authors thank Daniel Eisenstein, Julian Mu\~noz, Ramesh Narayan, Jacob Barandes, David Kaiser, and Robert Wald for insightful comments and discussions. We are deeply grateful to Jorma Louko for giving feedback on a draft of this paper {\red and to the anonymous journal referee for constructive comments and suggestions}.
  C.J.S. acknowledges support from the Quad Fellowship.

\vspace{-1em}

\section*{Data Availability}

The data that support the findings of this article are available at \url{https://doi.org/10.5281/zenodo.15522574}.

\vspace{-1em}

\appendix

\section{Comparison to Ng \textit{et al.}}\label{appendix:extra-plots}

\begin{figure*}[t]
  \includegraphics{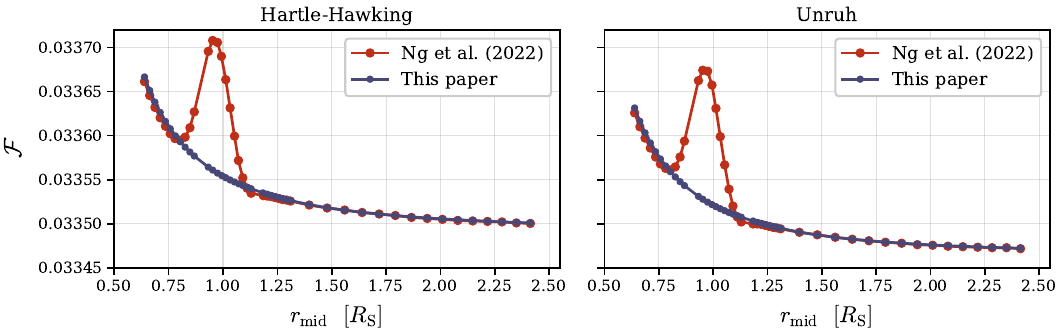}
  \caption{Comparison of our implementation to the results reported in Ref.~\cite{ng_LittleExcitementHorizon_2022} for a detector freely falling from rest at infinity into a Schwarzschild black hole, using $E = 5 / \RS \approx 63 \, \THawk$ and $\Delta\tau = 0.3 \, \RS \approx 0.024 / \THawk$.
  }\label{fig:schwarz-Ng-Fig1}
\end{figure*}

Figure~\ref{fig:schwarz-Ng-Fig1} compares our implementation to the results reported in Ref.~\cite{ng_LittleExcitementHorizon_2022}.

\bibliography{hawkrad-infalling}

\end{document}